\documentclass[a4paper,12pt]{article}
\pdfoutput=1
\usepackage[utf8]{inputenc}
\usepackage{amsmath}
\usepackage{amsfonts}
\usepackage{amssymb}
\usepackage{graphicx, rotating}
\usepackage{epsfig}
\usepackage{latexsym}
\usepackage{color}
\usepackage{bm}
\usepackage[colorlinks,allcolors=blue]{hyperref}
\usepackage[numbers,sort&compress]{natbib}

\usepackage[english]{babel}

\newcommand{\Slash}[1]{{\ooalign{\hfil#1\hfil\crcr\raise.167ex\hbox{/}}}}

\newcommand{\beq}{\begin{equation}}  \newcommand{\eeq}{\end{equation}}
\newcommand{\bef}{\begin{figure}}  \newcommand{\eef}{\end{figure}}
\newcommand{\bec}{\begin{center}}  \newcommand{\eec}{\end{center}}
  
\newcommand{\laq}[1]{\label{eq:#1}}  

\newcommand{\Eq}[1]{Eq.~(\ref{eq:#1})}

\newcommand{\Sec}[1]{Sec.\ref{chap:#1}}

\newcommand{\lac}[1]{\label{chap:#1}}

\def\({\left(}
\def\){\right)}

\def\O{\mathcal{O}}

\newcommand{\AND}{~{\rm and}~}
\newcommand{\EV}{ {\rm \, eV} }
\newcommand{\KEV}{ {\rm \, keV} }
\newcommand{\MEV}{ {\rm \, MeV} }
\newcommand{\GEV}{ {\rm \, GeV} }

\def\a{\alpha}

\def\f{\phi}
\def\g{\gamma}

\def\n{\nu}

\def\s{\sigma}

\def\D{\Delta}
\def\G{\Gamma}
\def\L{\Lambda}
\def\F{\Phi}

\def\*{\dagger}

\begin{document}
\renewcommand\bibname{\Large References}

\begin{flushright}
{TU-12XX}
\end{flushright}

\begin{center}

\vspace{0.1cm}

{\Large\bf {Explaining the KM3-230213A Detection without Gamma-Ray Emission: Cosmic-Ray Dark Radiation}}
\vspace{2cm}

{\bf Yuma Narita}$^{1}$ ~~~and~~~ {\bf   Wen Yin}$^{2}$\\
  \vspace{0.5cm}

 {\em 
(1) Department of Physics, Tohoku University, \\Sendai, Miyagi 980-8578, Japan} \\
 {\em 
(2) Department of Physics, Tokyo Metropolitan University,\\
 Hachioji-shi, Tokyo 192-0397 Japan} 

\abstract{
Recently, a high-energy neutrino event, designated KM3-230213A, was observed by the KM3NeT/ARCA detector in the Mediterranean Sea. This event is characterized by a reconstructed muon energy of approximately 120 PeV, corresponding to a median neutrino energy of roughly 220 PeV. The inferred average flux is estimated as 
$
E^2\Phi_{\nu} \simeq 5.8\times10^{-8}\,\mathrm{GeV\, cm^{-2}\, s^{-1}\, sr^{-1}}.
$
To understand the origin, it is essential to investigate consistency with multi-messenger observations—particularly gamma-ray constraints—in various theoretical scenarios within and beyond the Standard Model. Motivated by this, we explore the possibility that the detected event does not originate from conventional neutrinos but rather from right-handed neutrinos (sterile neutrinos) mixing with active neutrinos, leading to the observed muon signal. Such cosmic-ray dark radiation may have originated either in the early Universe or through dark matter decay in the present epoch. We show that in both cases, while satisfying existing general constraints on light sterile neutrinos, the stringent multi-messenger gamma-ray limits can be significantly alleviated. A distinct prediction of this scenario is that such events can arrive from directions through Earth that would typically attenuate conventional neutrinos. A related scenario involving cosmic-ray boosted WIMP dark matter is also discussed.

}

\end{center}
\clearpage
\section{Introduction}

Recently, the detection of ultra‐high‐energy cosmic neutrinos has opened new avenues for exploring the most extreme astrophysical phenomena. A landmark observation was reported by the KM3NeT Collaboration~\cite{KM3NeT:2025npi}, which detected an event designated KM3-230213A, with an estimated energy of approximately $220$ PeV, making it the highest-energy neutrino event recorded to date. The averaged neutrino antineutrino flux per each flavor, assuming a steady isotropic $\F_{\nu}\propto E^{-2}$ spectrum, is estimated to be 
$
E^2\Phi_{\nu} \simeq 5.8^{+10.1}_{-3.7}\times10^{-8}\,\mathrm{GeV\, cm^{-2}\, s^{-1}\, sr^{-1}}.
$

Either if neutrinos are produced solely within the Standard Model framework or not, additional multi-messenger signals are important to understand the data (c.f.~\cite{KM3NeT:2025aps, KM3NeT:2025vut}). In a beyond-Standard Model scenario—such as a heavy particle decay producing an energetic neutrino (e.g.,~\cite{Borah:2025igh,Kohri:2025bsn} for heavy dark matter decay)—one must also consider the decay into a charged lepton and charged particle(s) due to gauge invariance, which happens at the same probability and results in significant photon production if the mother particle is a gauge singlet.\footnote{An interesting loophole is that the decaying heavy particle is not a gauge singlet. Note that at high mass scales the direct detection constraints on the dark matter charge assignment are highly alleviated.} These photons are in tension with gamma-ray constraints from experiments such as KASCADE, KASCADE-Grande~\cite{KASCADEGrande:2017vwf}, and the Pierre Auger Observatory~\cite{Savina:2021cva, PierreAuger:2022uwd, PierreAuger:2022aty}, and LHAASO~\cite{LHAASO:2023gne} setting an upper bound on the photon flux 
$
E^2\Phi_\g \lesssim 10^{-10} - 10^{-9}\GEV \rm cm^{-2} s^{-1} sr^{-1},
$ in a similar energy range (e.g.,~\cite{LHAASO:2022yxw,Das:2023wtk} for the limit of decaying dark matter scenario).

However, the detection of this event is in tension~\cite{KM3NeT:2025ccp} with the IceCube data~\cite{IceCube:2018fhm,IceCube:2020wum} and the Auger data~\cite{AbdulHalim:2023SN}, even when assuming a point source~\cite{IceCube:2018fhm,KM3NeT:2018wnd,Li:2025tqf}. This tension can be alleviated if, under specific Earth matter effects, sterile neutrinos from a point source oscillate into detectable neutrinos~\cite{Brdar:2025azm}. Although the tension between the IceCube and KM3NeT data calls for a better understanding of the event itself, this may be an opportune time to study a novel mechanism that suppresses multi-messenger signals while still producing a significant event rate in neutrino detectors. In this paper, we illustrate such a mechanism by interpreting the KM3NeT event.

We consider the possibility that {\it the detected event originates from an accelerated right-handed neutrino} (RHN, or sterile neutrino), $N$. It couples to Standard Model particles in the same manner as active neutrinos but with a coupling suppressed by a small mixing parameter $\theta(\ll 1)$. In the broken phase of electroweak symmetry, the mass term is given by (see \cite{Abazajian:2012ys,Boyarsky:2009ix,Drewes:2016upu,Abazajian:2017tcc,Acero:2022wqg} for the review of the model)
\beq
{\cal L} = -\theta m_N \bar{\nu}_R \nu_\a - \frac{m_N}{2} \bar{\nu}_R^{c} \nu_R + {\rm h.c.}
\eeq
where $\a$ denotes the neutrino flavor, $\nu_\a$ is the left-handed neutrino, and $m_N$ is the mass of the RHN. By diagonalizing the mass matrix, we can readily see that the heavy eigenstate with mass $\simeq m_N$ is 
$$
N \simeq \theta \nu_\a + \nu_R.
$$
This explains why $N$ has a similar, yet suppressed, coupling compared to $\nu_\a$. For simplicity, hereafter we consider mixing with a single flavor and omit the index $\a$ unless otherwise stated.

Neglecting the RHN and neutrino masses, the scattering cross section for a process involving a single RHN and any Standard Model particle is given by \beq
\s_{N}(E_{\rm cm}) \simeq \theta^2 \s_{\n}(E_{\rm cm}),
\eeq
for $\theta \ll 1$, where $\sigma_\nu$ is the crosssection for  replacing $N$ by the the usual neutrino. 
Therefore, if the mass $m_N$ is negligible, which is the case of the high energy event, the neutrino event can be mimicked by the RHN cosmic ray if the flux $\F_N$ satisfies
\beq
\laq{FN}
\boxed{\F_{N }(E) \simeq C^{-1} \theta^{-2} \F_{\nu}(E).}
\eeq
Here, $\F_\nu$ denotes the flux of standard neutrinos, and we assumed the total number of the scatterer in the water of Mare Mediteraneum is the same for the RHN and the neutrino~(c.f.~\cite{Cherry:2018rxj,Huang:2018als}). 
Neutrinos with energies around $10^{8\text{-}9}\,\GEV$ can penetrate Earth only if they arrive close to the horizon, whereas RHNs with sufficiently small mixing angle $\theta$ can traverse Earth from other directions (see Ref.\,\cite{Li:2025tqf} for details on the neutrino penetration solid angle) consistent with other observation. $C = \O(1-10)$ is the enhancement factor arising from differences in effective area.\footnote{This factor also incorporates uncertainties arising from approximations, spectral dependence, and flavor dependence of the relevant reactions.} 

A similar scenario was previously studied in Ref.\,\cite{Jaeckel:2020oet} in the context of probing the reheating phase. In that work, the authors examined the parameter region where $N$ is produced by reheating prior to Big Bang Nucleosynthesis (BBN) and eventually arrives at Earth. The absence of thermalization of the primordial cosmic ray significantly restricts the allowed parameter region. Thus, their main focus was on RHN dark radiation decaying primarily into three neutrinos. Although this decay can also produce charged particles and other radiation, it occurs well before the present epoch, thereby evading cosmic-ray constraints from Earth-based observations. Furthermore, there exist parameter regions that evade constraints from the cosmic microwave background (CMB) and BBN. Indeed, the KM3NeT event could be explained within such a framework if one allows a super-Planckian mass for the reheating particle (see Fig.~4 of Ref.\,\cite{Jaeckel:2020oet}).  

Later, in Ref.\,\cite{Jaeckel:2021ert}, the scenario involving axion-like particle radiation produced by modulus decay—including epochs after BBN—was discussed, highlighting the significance of (relativistic and non-relativistic) dark radiation decays into photons, and the usefulness of X-ray and $\gamma$-ray observations as potential probes.

In the present paper, we specifically study RHN cosmic-ray mimicing the neutrino event without requiring that $N$ is produced exclusively before BBN. This will provide a new interpretation of high energy neutrinos, and the future prospects are very different from the previous studies.

\begin{figure}[t!]
    \begin{center}
      \includegraphics[width = 150mm]{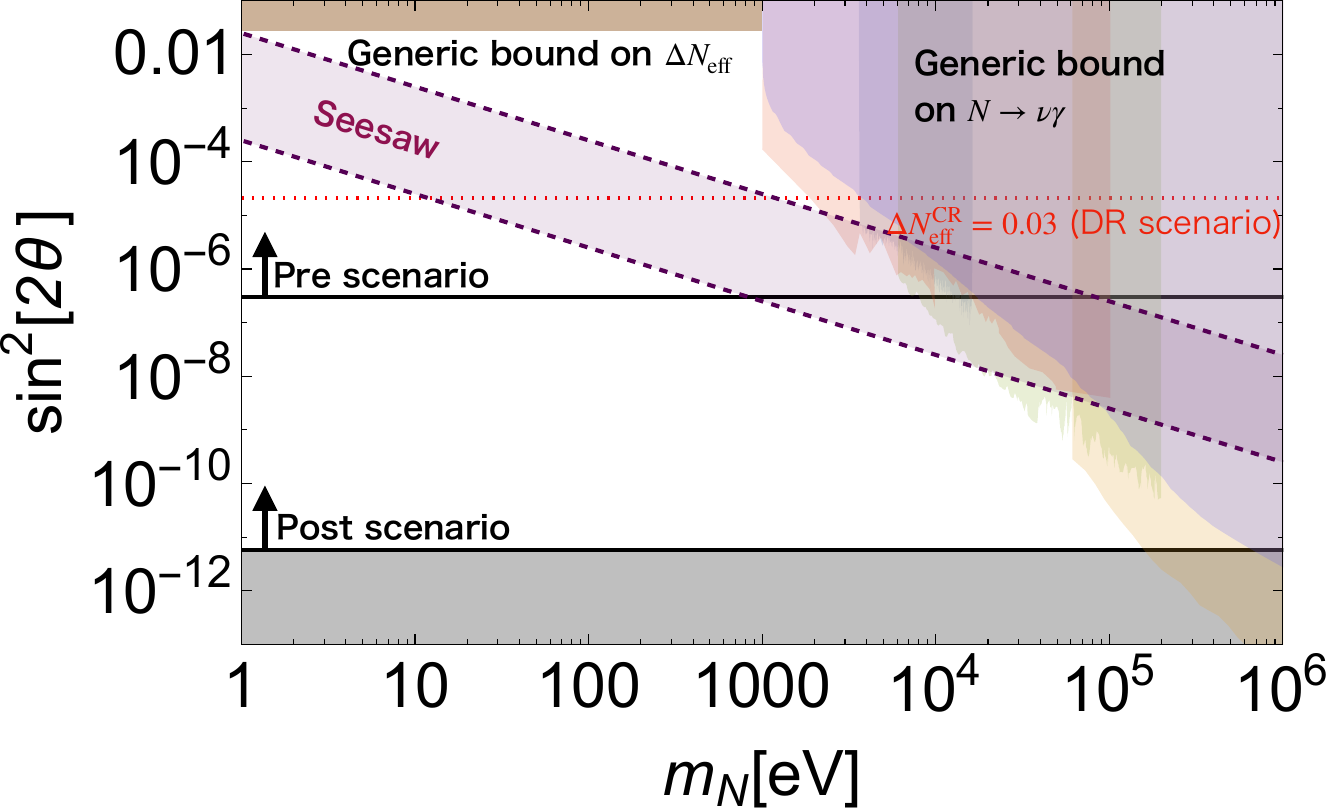}  
    \end{center}
  \caption{Allowed region in $m_N$ and $\sin^2(2\theta)$ for the RHN as the cosmic-ray dark radiation to explain the KM3NeT event.  
The upper colorfully shaded regions are excluded due to RNH produced thermally with $T_R=1.8\MEV$ includes the BBN bound on the $\D N_{\rm eff}$ and the bound on $N\to \nu \g$. The purple shaded region, bounded by the dashed lines, represents the seesaw relation. The upper horizontal solid line denotes the limit from the CMB bound on the $\D N_{\rm eff}$ for the pre-recombination scenario. Below the red dotted line, denoting the contour for $\D N_{\rm eff}\approx 0.03$, the parameter region can be probed by future measurements of $\D N_{\rm eff}$. 
The lower solid line denotes the limit on the post-recombination scenario assuming the dark matter decay to produce the dark radiation.  We take $C=7, \AND d_\a=0.79$. }
  \label{fig:1}
\end{figure}

Depending on the origin of the RHN, we consider two scenarios. In one scenario, the RHN radiation is generated before the recombination era, which we refer to as the {\it pre recombination scenario}. In the other scenario, the RHN is generated after recombination; in this case, we consider production via the decay of scalar particle into a pair of $N$, {\it post recombination scenario}. We demonstrate that the scenario allows for a large parameter region, as shown in Fig.~\ref{fig:1}, in terms of the RHN mass $m_N$ and $\sin^2 (2\theta)$. In this figure, the top and right top colored shaded regions are excluded model-independently due to a thermal production of $N$, which will be discussed in \Sec{2}.  The purple band delineated by the dashed line representing the seesaw relation $\theta^2 m_N = 0.001\EV - 0.1\EV$~\cite{Minkowski:1977sc,Yanagida:1979as,Glashow:1979nm, GellMann:1980vs,Mohapatra:1979ia}, explaining the neutrino mass.\footnote{Above this band, an approximate lepton number symmetry is required to suppress the induced neutrino mass. Below the band only the lightest neutrino mass can be explained.} In the regions above the two solid black lines, the pre-recombination scenario (top) and the post-recombination scenario (whole region) which will be discussed in \Sec{3} can explain the KM3Net event.
In these allowed region, the constraint from the high energy $\gamma$-ray bound is highly alleviated (see~\Sec{4}).
The last section is devoted to the conclusions and discussion.

\section{Generic Bound on Sterile Neutrino}
\lac{2} 
Before discussing the details of cosmic-ray dark radiation, let us derive a model-independent bound on the thermally produced sterile neutrino. A similar bound is derived in \cite{Langhoff:2022bij} for sterile neutrino mixing with electron neutrino using the X-ray and $\g$-ray limit on the dark radiation \cite{Jaeckel:2021ert} (see also \cite{Cadamuro:2011fd}). 
We do not concentrate on the electron mixing, and study the limit imposed mostly by the photon constraint for $N \to \nu \gamma$ reaction by using various new data.  
The (sub MeV) RHN radiative decay rate is given by \cite{Pal:1981rm,Boyarsky:2018tvu}, 
\beq
\G_N^{-1}\approx 2.3\times 10^{24} {\rm yr} \frac{10^{-10}}{\sin^2(2\theta)} \(\frac{\KEV}{m_N}\)^5,
\eeq
which is flavor independent.

In the low reheating temperature scenario, which will yield the most conservative bound, the abundance of the RHN is estimated by~\cite{Gelmini:2004ah} (c.f. \cite{Dodelson:1993je})
\beq
\Omega_N h^2 \simeq 0.1 \, d_{\alpha} \left( \frac{\sin^2 2\theta}{10^{-3}} \right) \left( \frac{m_N}{1 \,\text{keV}} \right) \left( \frac{T_R}{5 \,\text{MeV}} \right)^3
\eeq
where $d_\a=1.13, \AND 0.79$ are for $\a=e$ and $\mu,\tau$ respectively. $h$ is the reduced Hubble parameter. $T_R$ is the reheating temperature. 
 We take $T_R=1.8\MEV$ as the lowest value allowed by the BBN bound~\cite{Hasegawa:2019jsa} (see also Refs.~\cite{Kawasaki:1999na,Kawasaki:2000en,  Hannestad:2004px, Ichikawa:2006vm, DeBernardis:2008zz, deSalas:2015glj, Hasegawa:2019jsa}).
 Then, we obtain the smallest RHN abundance produced in  the Big-Bang cosmology for  a given set of $\theta^2\AND m_N$. 

The typical momentum of the thermally produced RHN can be estimated by \beq
\bar{p}_N= \frac{7 \pi ^4}{180 \zeta (3)} T_{\nu}
\eeq
assuming that the distribution is similar to that of neutrinos. Here $T_\nu= (4/11)^{1/3} T_{\g}$ is the neutrino temperature with the photon temperature, $T_\g$. 

For $m_N\gtrsim \KEV$, the typical velocity at matter-radiation equality is given by $\bar p_\nu/m_N\lesssim 5\times 10^{-4}$, which is so small that we expect these particles to be captured by galaxies and galaxy clusters since the escape velocity is $\gtrsim 10^{-3}$. The distributions of $N$ in the galaxies and galaxy clusters follow the dominant dark matter distributions if they reach Jeans equilibrium, but $N$ constitutes a fraction $r\equiv \frac{\Omega_N h^2}{0.12}$ of the total dark matter. Using this fact, we can recast various dark matter constraints from the X-ray and $\gamma$-ray observations including XMM-Newton~\cite{Gewering-Peine:2016yoj, Foster:2021ngm}, combined NuSTAR~\cite{Perez:2016tcq,Ng:2019gch,Roach:2022lgo}, INTEGRAL~\cite{Calore:2022pks}, XRISM ~\cite{Yin:2025xad}, and COMPTEL, EGRET, Fermi Gamma-ray Space Telescope \cite{Essig:2013goa,Boyarsky:2018tvu}. 
Also shown is the cosmic background bound~\cite{Porras-Bedmar:2024uql}. 
With much larger $\sin^2(2\theta)$, the proper lifetime of $N$ is shorter than the age of the Universe. Nevertheless, serious constraints remain from the cosmic backgrounds as well as from BBN. We do not distinguish them.  These limits, taken from \cite{AxionLimits} and~\cite{Yin:2025xad} for the limit from XRISM, are shown in the top-right colorfully shaded region.\footnote{The mechanism presented in Ref.\,\cite{Brdar:2025azm} does not work in our scenario either due to this bound and that the dark radiation is not from the point source. In some extensions of the model, e.g., by adding a light modulus to alter the mass of the RHN in the early Universe, the production of RHNs can be suppressed, thereby alleviating the limit derived here~\cite{Yin:2024trc}.} 

Since the mass is typically small in the large mixing limit due to these constraints, the RHN behaves as dark radiation during the BBN era. 
The amount in terms of the deviation of the effective neutrino number is $\D N_{\rm eff}\sim \Omega_N \frac{\bar{p}^{(0)}_N}{m_N} \rho_c/(7/4\times \pi^2/30\times (T_\nu^{(0)})^4)$. Here and hereafter, the superscript $``(0)"$ denotes the value measured in the present Universe. 
 Then we have a constraint of $\D N_{\rm eff}\lesssim 0.2$~\cite{Cyburt:2015mya}, which is shown in the top shaded region. This bound is also generic.

The derived limits are almost flavor independent by neglecting the small dependence of $d_\a$. 
Depending on the neutrino flavor, we may have more stringent limits such as the supernova bound~\cite{Shi:1993ee,Suliga:2020vpz} and bound from particle experiments.

\section{Fake Neutrino Flux and the Lower Bounds of the Mixing Parameter}
\lac{3}
Now we are ready to discuss the scenario to explain the KM3-230213A event.

\paragraph{Pre-Recombination Scenario}

In various  cosmological scenarios, dark radiation in the form of right-handed neutrinos, $N$, can be naturally produced. For instance, a real scalar, $\f$, which may be a modulus, decays into a pair of $N$ after the BBN (see \cite{Jaeckel:2020oet}). 
Indeed, one can always include a term of the form
\beq
\laq{lag2}
{\cal L} \supset \f \bar{N^c} N.
\eeq
Since the decay completes before the recombination for the pre-recombination scenario, we assume $\f$ is always a subdominant component the Universe. 

Alternatively, a strongly first-order phase transition in a dark sector is also plausible, since a heavy particle can be produced by the ``scattering'' between the Higgs field in an ultra-relativistic bubble wall and the ambient plasma~\cite{Azatov:2020ufh,Azatov:2021ifm,Baldes:2022oev,Azatov:2024crd,Ai:2024ikj}. The spectra can be found in \cite{Azatov:2024crd} for the production via higher dimensional term. 

Irrespective of the details of the scenario, we can parameterize the dark radiation component by the deviation of the effective neutrino number for this primordial cosmic-ray dark radiation, $\D N^{\rm CR}_{\rm eff}$. The energy density at any cosmic time after recombination (or after production ceases) is given by
$
\rho_N = \D N^{\rm CR}_{\rm eff}  \left(4/11\right)^{4/3} 7/8 \times \rho_\g.
$
This leads to an energy flux satisfying
\beq
\int dE\, E \F_N(E) \sim 0.14 \D N^{\rm CR}_{\rm eff} \,\mathrm{GeV\,cm^{-2}\,s^{-1}\,sr^{-1}}.
\eeq
Assuming that the spectrum is localized around the peak position, we can approximate the left-hand side by $E^2 \F_N(E)$. Then, using \Eq{FN}, we obtain
\beq
\D N^{\rm CR}_{\rm eff} \sim 1.6\times 10^{-6} C^{-1} \theta^{-2} \frac{E^2 \F_\nu(E)}{5.8\times 10^{-8} \mathrm{GeV\,cm^{-2}\,s^{-1}\,sr^{-1}}}.
\eeq
Immediately, we get a lower bound on the mixing angle from the CMB constraint, $\D N^{\rm CR}_{\rm eff} \lesssim 0.28$ (2$\s$ limit)~\cite{Planck:2018vyg}:
\beq
\theta^2 \gtrsim 2.1\times 10^{-6} C^{-1},
\eeq
where we have used the lowest value of $E^2 \F_\nu(E) \approx 2.1\times 10^{-8}\,\mathrm{GeV\,cm^{-2}\,s^{-1}\,sr^{-1}}$ within the $1\s$ error, corresponding to the upper horizontal black solid line in Fig.~\ref{fig:1}. The red dashed line can be derived by adopting $\D N^{\rm CR}_{\rm eff} =0.03$, which is the sensitivity reach of CMB-S4~\cite{CMB-S4:2016ple}.

\paragraph{Post-Recombination Scenario} 

\begin{figure}[t!]
    \begin{center}
      \includegraphics[width = 150mm]{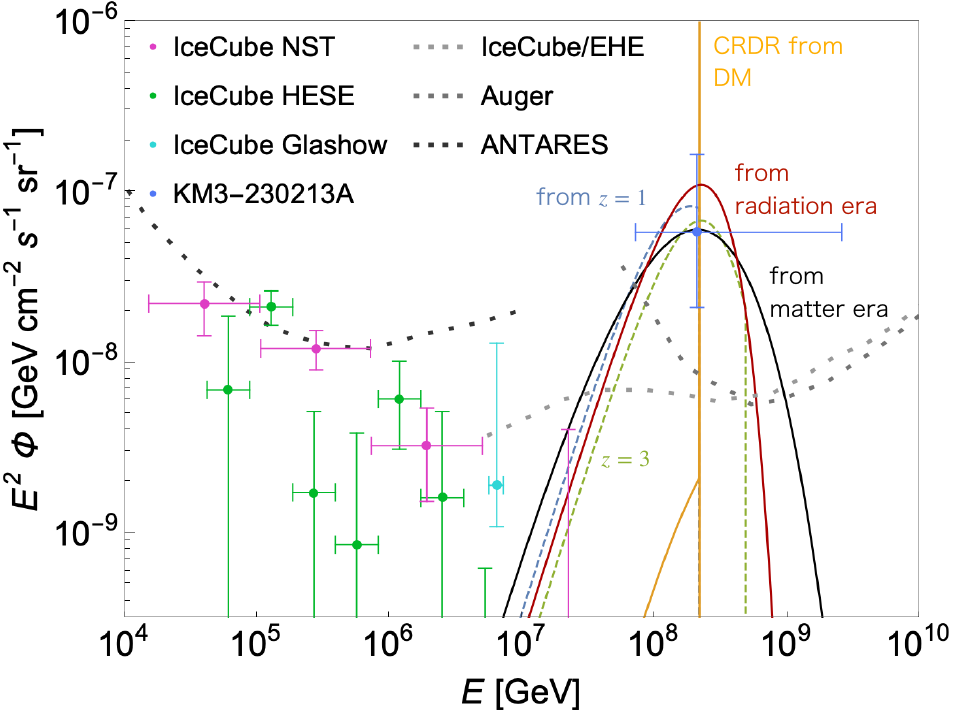}  
    \end{center}
  \caption{The fake neutrino spectrum, $C \theta^2 E^2 \Phi_N$, originates from the two-body decay of a non-relativistic real scalar $\phi$.  In the post-recombination scenario, cases considered include $\phi$ as decaying dark matter (orange solid line), as a modulus decaying completely during the matter-dominated era (radiation solid line), and as decaying incompletely at redshifts $z=1$ (blue dashed line) and $z=3$ (green dashed line). The post-recombination scenario also includes decay during the radiation-dominated era (red solid line). The spectral height and energy are chosen appropriately. The data points and constraints are taken from \cite{KM3NeT:2025ccp}. These data are shown as NST (purple points)~\cite{Abbasi:2021qfz}, HESE (green points)~\cite{IceCube:2020wum}, and Glashow resonance event(light blue points)~\cite{IceCube:2021rpz}. Also, the dotted gray lines represent upper limits from IceCube-EHE (90\% CL~\cite{IceCube:2018fhm}), Auger (90\% CL~\cite{PierreAuger:2023pjg}), and ANTARES (90\% CL~\cite{ANTARES:2024ihw}).}
  \label{fig:2}
\end{figure}

For the post-recombination scenario, let us first consider that $\f$ represents the dominant dark matter particle with the same Lagrangian \Eq{lag2}.
 The $N$ flux from the galactic dark matter decay can be estimated from
\beq
\int dE\F_N(E) \approx \G_\f \frac{D}{2\pi m_\f},
\eeq
where $D \equiv \int d\Omega\, ds\, \rho_\phi$ is the $D$ factor, and $\rho_\phi$ is the dark matter density profile. Here, we adopt the Navarro--Frenk--White (NFW) profile~\cite{Navarro:1996gj}:
$
\rho_\phi(r) = r_{s}r^{-1}\rho_{0}\left(r/r_{s}+1\right)^{-2},
$
with $r$ being the distance from the galactic center. The values fitted from Gaia DR2~\cite{Cautun:2019eaf} are $\rho_{0}=0.46\,\GEV\,\rm cm^{-3}$ and $r_s=14.4$ kpc, and we use
$
r = \sqrt{s^2+R^2_{\odot}-2R_{\odot}s \cos{\Theta}},
$
with $R_{\odot}=8.2$ kpc and $\Theta$ being the angle between the line of sight and the galactic center. This gives $D \approx 2.2\times 10^{23} \GEV\rm cm^{-2}$. An extragalactic component is also present, but it is suppressed compared to the galactic one (see \Sec{4} and Fig.~\ref{fig:2}). Since the galactic flux is monochromatic while the analysis in \cite{KM3NeT:2025npi} assumes an $E^{-2}$ scaling spectrum, a more detailed fit is necessary for a precise estimation, which is beyond our scope. As a rough approximation, we use
\beq
E^2 \F_N(E) \sim \frac{m_\f}{2}\times \int dE\F_N(E)
\eeq
to mimic the neutrino flux via the relation in \Eq{FN}. Then we obtain
\beq
\G_\f^{-1} \sim C \theta^2 \times 2.9^{+4.7}_{-1.8} \times 10^{29} \rm s.
\eeq

Since the decay into dark particles is constrained 
$
\G_{\f} < (246\rm Gyr)^{-1},
$
(e.g.,~\cite{Enqvist:2019tsa,Alvi:2022aam}), we obtain a lower bound on the mixing:
\beq
\theta^2 \gtrsim C^{-1} 1.0\times 10^{-11},
\eeq
where we again have used the upper limit for $\G^{-1}_\f$ to set a conservative limit for explaining the event. This bound, shown as the lower horizontal solid line in Fig.~\ref{fig:1}, is weaker than the pre-recombination case because, during the matter- and dark energy-dominated epochs, radiation production can be more efficient without affecting the overall cosmic history. However, in this scenario, we do not obtain a bound from $\D N^{\rm CR}_{\rm eff}$ because, at the time of recombination, the production of $N$ is not as significant as it is today (see, e.g., Fig. 3 of Ref.\,\cite{Batell:2024hzo}).

Alternatively it is also possible that 
the production of the cosmic-ray dark radiation is before today but after recombination. 
For instance, a complete or almost complete decay of the real scalar field can produce it. 
Usually the allowed region is smaller than the case of dark matter because the total energy density of $\phi$ and thus the resulting comoving energy density of $N$ is restricted.

The various fake neutrino spectrum from $\phi$ decay  are shown in Fig.\ref{fig:2}. 
In the post-recombination scenario, cases considered include $\phi$ as decaying dark matter (orange solid line), as a modulus decaying completely during the matter-dominated era (black solid line), and as decaying incompletely at redshifts $z=1$ (blue dashed line) and $z=3$ (green dashed line). The post-recombination scenario also includes decay during the radiation-dominated era (red solid line). The spectral height and energy are chosen appropriately.
They fit the data well.

\section{Alleviated $\gamma$ Ray Bounds from the Cosmic-Ray Dark Radiation}
\lac{4}
Given the previous constraints, we show that the severe multi-messenger limits on the cosmic-ray dark radiation are absent. 

\paragraph{Pre-Recombination Scenario}
Since $N$ is boosted to an energy of $E$, the decay rate is suppressed by a factor of $\G_N m_N/E$, which redshifts with $(1+z)$, where $z$ is the redshift. Then the energy density of the photon can be estimated 
$
\rho^{(0)}_{\g}\approx \int d z \frac{\G_N m_N}{E_N^{(0)}H}\rho^{(0)}_N \approx 3.2\frac{\G_N m_N}{E_N^{(0)}H^{(0)}}\rho^{(0)}_N.
$
Here, $H=H^{(0)}\sqrt{(1+z)^4 \Omega_r+ (1+z)^3 \Omega_m +\Omega_{\L}}$ is the Hubble parameter, taking into account the contributions from radiation, matter, and dark energy: $\Omega_r\approx 9.2 \times10^{-5}, \Omega_m \approx 0.32, \Omega_{\L}\approx 0.68$, and $H_0\approx 67.8$ km/s/Mpc. 
Note that the production is most efficient in the late Universe and thus the photon energy is close to the $N$ energy. 
Here $E_N^{(0)}\approx 220\,$PeV, and $ \rho_N^{(0)}/(4\pi)\approx \int d E E C^{-1}\theta^{-2}\F_\nu(E)\sim
 C^{-1}\theta^{-2}E^2 \F_\nu(E) \sim
 C^{-1}\theta^{-2} 5.8\times 10^{-8}\GEV \rm cm^{-2}s^{-1} sr^{-1}$ for explaining the event. 
By requiring the photon flux $E^2\F_\g\lesssim 10^{-9} \GEV \rm cm^{-2}s^{-1} sr^{-1}$, which is the typical order of magnitude for the photon limits in a similar mass range~\cite{KASCADEGrande:2017vwf,Savina:2021cva, PierreAuger:2022uwd, PierreAuger:2022aty}, we obtain the bound \beq 
\laq{multi}
m_N\lesssim 0.4 \MEV C^{1/6}\eeq
which is independent of $\theta$. 
These bounds are automatically satisfied in the parameter region in Fig.\ref{fig:1}. 
In the allowed parameter region, the dominant decay of $N\to \nu \bar \nu \nu$, which may contribute to the neutrino spectrum, is also suppressed.

\paragraph{Post-Recombination Scenario}
For the scenario in which an $N$ pair is produced from the decay of heavy dark matter in the galactic halo, $\f \to N N$, the further decay of $N$ to $\nu \g$ is suppressed by the length scale of the galaxy. 
The photon flux is predominantly obtained from the extra galactic component (see, e.g., Ref.,\cite{Ho:2019ayl} for the analytic formula). We obtain  $E_N^2 \F_N^{\rm ext}(E_N)|_{E_N\approx m_\f/2}\approx C^{-1} \theta^{-2} 2.1\times 10^{-9} \GEV\rm cm^{-2}s^{-1}sr^{-1}$ for explaining the event. The same procedure as before yields a similar constraint, $m_N\lesssim 0.1 \MEV C^{1/6}$, as given in \Eq{multi}. The constraint from other scenarios are also similar. 

Therefore, we conclude that our scenario is free from high energy $\gamma$-ray bounds and, trivially, from bounds on the nucleon--antinucleon cosmic-rays. 

\section{Conclusions and Discussion}
We have shown that if the RHN dark radiation mimics the neutrino event, the constraints from multi-messenger observation can be highly alleviated. In other words, neutrino detectors such as IceCube, KM3NeT, and others can serve as RHN detectors.  
If this cosmic-ray dark radiation scenario is correct, the smoking-gun signal would be a high-energy neutrino event arriving from an unusual direction--one through which ordinary neutrinos cannot normally pass. More high energy events are needed to confirm this possibility.

In fact, if this is indeed due to an RHN, we obtain a very valuable messenger of the early Universe~\cite{Jaeckel:2020oet}. Given that the RHN can be easily produced from the two-body decay of a heavy scalar particle, the spectrum is monochromatic in the rest frame of the mother particle. If the mother particle is non-relativistic, the $N$ spectrum depends on the expansion history via redshift (see Fig.\ref{fig:2}). In particular, precise measurement of the spectrum can distinguish among scenarios for non-relativistic $\phi$ decay (see Fig.\ref{fig:2}), or arises from  other sources~\cite{Jaeckel:2021gah}.

Another similar scenario  involves cosmic-ray--boosted light WIMP dark matter via cosmogenic neutrino interacting with the cosmic-neutrino background~\cite{Yin:2018yjn} (see also later studies~\cite{Bringmann:2018cvk,Ema:2018bih,Cappiello:2018hsu}). In this scenario, the “high-energy cosmic ray” is in fact dark matter rather than a cosmogenic neutrino, which would otherwise be attenuated by the new interaction. This boosted dark matter can further interact with the detector—mimicking a neutrino—or it can interact with the Earth or atmosphere to produce high-energy neutrinos that eventually reach the detector. It is worth investigating whether this scenario could help alleviate the tension between KM3NeT and IceCube.

\section*{Acknowledgement}
We would like to thank K. Murase for useful discussions in a different project. 
This work is supported by JSPS KAKENHI Grant Nos.  20H05851 (W.Y.),  22K14029 (W.Y.), 22H01215 (W.Y.), Graduate Program on Physics for the Universe (Y.N.), and JST SPRING, Grant Number JPMJSP2114 (Y.N.). W.Y. is also supported by Incentive Research Fund for Young Researchers from Tokyo Metropolitan University.

\bibliographystyle{apsrev4-1}
\bibliography{KM3NET2.bib}

\begin{thebibliography}{82}%
\makeatletter
\providecommand \@ifxundefined [1]{%
 \@ifx{#1\undefined}
}%
\providecommand \@ifnum [1]{%
 \ifnum #1\expandafter \@firstoftwo
 \else \expandafter \@secondoftwo
 \fi
}%
\providecommand \@ifx [1]{%
 \ifx #1\expandafter \@firstoftwo
 \else \expandafter \@secondoftwo
 \fi
}%
\providecommand \natexlab [1]{#1}%
\providecommand \enquote  [1]{``#1''}%
\providecommand \bibnamefont  [1]{#1}%
\providecommand \bibfnamefont [1]{#1}%
\providecommand \citenamefont [1]{#1}%
\providecommand \href@noop [0]{\@secondoftwo}%
\providecommand \href [0]{\begingroup \@sanitize@url \@href}%
\providecommand \@href[1]{\@@startlink{#1}\@@href}%
\providecommand \@@href[1]{\endgroup#1\@@endlink}%
\providecommand \@sanitize@url [0]{\catcode `\\12\catcode `\$12\catcode
  `\&12\catcode `\#12\catcode `\^12\catcode `\_12\catcode `\%12\relax}%
\providecommand \@@startlink[1]{}%
\providecommand \@@endlink[0]{}%
\providecommand \url  [0]{\begingroup\@sanitize@url \@url }%
\providecommand \@url [1]{\endgroup\@href {#1}{\urlprefix }}%
\providecommand \urlprefix  [0]{URL }%
\providecommand \Eprint [0]{\href }%
\providecommand \doibase [0]{http://dx.doi.org/}%
\providecommand \selectlanguage [0]{\@gobble}%
\providecommand \bibinfo  [0]{\@secondoftwo}%
\providecommand \bibfield  [0]{\@secondoftwo}%
\providecommand \translation [1]{[#1]}%
\providecommand \BibitemOpen [0]{}%
\providecommand \bibitemStop [0]{}%
\providecommand \bibitemNoStop [0]{.\EOS\space}%
\providecommand \EOS [0]{\spacefactor3000\relax}%
\providecommand \BibitemShut  [1]{\csname bibitem#1\endcsname}%
\let\auto@bib@innerbib\@empty
\bibitem [{\citenamefont {Aiello}\ \emph {et~al.}(2025)\citenamefont {Aiello}
  \emph {et~al.}}]{KM3NeT:2025npi}%
  \BibitemOpen
  \bibfield  {author} {\bibinfo {author} {\bibfnamefont {S.}~\bibnamefont
  {Aiello}} \emph {et~al.} (\bibinfo {collaboration} {KM3NeT}),\ }\href
  {\doibase 10.1038/s41586-024-08543-1} {\bibfield  {journal} {\bibinfo
  {journal} {Nature}\ }\textbf {\bibinfo {volume} {638}},\ \bibinfo {pages}
  {376} (\bibinfo {year} {2025})}\BibitemShut {NoStop}%
\bibitem [{\citenamefont {Adriani}\ \emph
  {et~al.}(2025{\natexlab{a}})\citenamefont {Adriani} \emph
  {et~al.}}]{KM3NeT:2025aps}%
  \BibitemOpen
  \bibfield  {author} {\bibinfo {author} {\bibfnamefont {O.}~\bibnamefont
  {Adriani}} \emph {et~al.} (\bibinfo {collaboration} {KM3NeT}),\ }\href@noop
  {} {\  (\bibinfo {year} {2025}{\natexlab{a}})},\ \Eprint
  {http://arxiv.org/abs/2502.08387} {arXiv:2502.08387 [astro-ph.HE]}
  \BibitemShut {NoStop}%
\bibitem [{\citenamefont {Adriani}\ \emph
  {et~al.}(2025{\natexlab{b}})\citenamefont {Adriani} \emph
  {et~al.}}]{KM3NeT:2025vut}%
  \BibitemOpen
  \bibfield  {author} {\bibinfo {author} {\bibfnamefont {O.}~\bibnamefont
  {Adriani}} \emph {et~al.} (\bibinfo {collaboration} {KM3NeT}),\ }\href@noop
  {} {\  (\bibinfo {year} {2025}{\natexlab{b}})},\ \Eprint
  {http://arxiv.org/abs/2502.08508} {arXiv:2502.08508 [astro-ph.HE]}
  \BibitemShut {NoStop}%
\bibitem [{\citenamefont {Borah}\ \emph {et~al.}(2025)\citenamefont {Borah},
  \citenamefont {Das}, \citenamefont {Okada},\ and\ \citenamefont
  {Sarmah}}]{Borah:2025igh}%
  \BibitemOpen
  \bibfield  {author} {\bibinfo {author} {\bibfnamefont {D.}~\bibnamefont
  {Borah}}, \bibinfo {author} {\bibfnamefont {N.}~\bibnamefont {Das}}, \bibinfo
  {author} {\bibfnamefont {N.}~\bibnamefont {Okada}}, \ and\ \bibinfo {author}
  {\bibfnamefont {P.}~\bibnamefont {Sarmah}},\ }\href@noop {} {\  (\bibinfo
  {year} {2025})},\ \Eprint {http://arxiv.org/abs/2503.00097} {arXiv:2503.00097
  [hep-ph]} \BibitemShut {NoStop}%
\bibitem [{\citenamefont {Kohri}\ \emph {et~al.}(2025)\citenamefont {Kohri},
  \citenamefont {Paul},\ and\ \citenamefont {Sahu}}]{Kohri:2025bsn}%
  \BibitemOpen
  \bibfield  {author} {\bibinfo {author} {\bibfnamefont {K.}~\bibnamefont
  {Kohri}}, \bibinfo {author} {\bibfnamefont {P.~K.}\ \bibnamefont {Paul}}, \
  and\ \bibinfo {author} {\bibfnamefont {N.}~\bibnamefont {Sahu}},\ }\href@noop
  {} {\  (\bibinfo {year} {2025})},\ \Eprint {http://arxiv.org/abs/2503.04464}
  {arXiv:2503.04464 [hep-ph]} \BibitemShut {NoStop}%
\bibitem [{\citenamefont {Apel}\ \emph {et~al.}(2017)\citenamefont {Apel} \emph
  {et~al.}}]{KASCADEGrande:2017vwf}%
  \BibitemOpen
  \bibfield  {author} {\bibinfo {author} {\bibfnamefont {W.~D.}\ \bibnamefont
  {Apel}} \emph {et~al.} (\bibinfo {collaboration} {KASCADE Grande}),\ }\href
  {\doibase 10.3847/1538-4357/aa8bb7} {\bibfield  {journal} {\bibinfo
  {journal} {Astrophys. J.}\ }\textbf {\bibinfo {volume} {848}},\ \bibinfo
  {pages} {1} (\bibinfo {year} {2017})},\ \Eprint
  {http://arxiv.org/abs/1710.02889} {arXiv:1710.02889 [astro-ph.HE]}
  \BibitemShut {NoStop}%
\bibitem [{\citenamefont {Savina}(2021)}]{Savina:2021cva}%
  \BibitemOpen
  \bibfield  {author} {\bibinfo {author} {\bibfnamefont {P.}~\bibnamefont
  {Savina}} (\bibinfo {collaboration} {Pierre Auger}),\ }\href {\doibase
  10.22323/1.395.0373} {\bibfield  {journal} {\bibinfo  {journal} {PoS}\
  }\textbf {\bibinfo {volume} {ICRC2021}},\ \bibinfo {pages} {373} (\bibinfo
  {year} {2021})}\BibitemShut {NoStop}%
\bibitem [{\citenamefont {Abreu}\ \emph {et~al.}(2022)\citenamefont {Abreu}
  \emph {et~al.}}]{PierreAuger:2022uwd}%
  \BibitemOpen
  \bibfield  {author} {\bibinfo {author} {\bibfnamefont {P.}~\bibnamefont
  {Abreu}} \emph {et~al.} (\bibinfo {collaboration} {Pierre Auger}),\ }\href
  {\doibase 10.3847/1538-4357/ac7393} {\bibfield  {journal} {\bibinfo
  {journal} {Astrophys. J.}\ }\textbf {\bibinfo {volume} {933}},\ \bibinfo
  {pages} {125} (\bibinfo {year} {2022})},\ \Eprint
  {http://arxiv.org/abs/2205.14864} {arXiv:2205.14864 [astro-ph.HE]}
  \BibitemShut {NoStop}%
\bibitem [{\citenamefont {Abreu}\ \emph {et~al.}(2023)\citenamefont {Abreu}
  \emph {et~al.}}]{PierreAuger:2022aty}%
  \BibitemOpen
  \bibfield  {author} {\bibinfo {author} {\bibfnamefont {P.}~\bibnamefont
  {Abreu}} \emph {et~al.} (\bibinfo {collaboration} {Pierre Auger}),\ }\href
  {\doibase 10.1088/1475-7516/2023/05/021} {\bibfield  {journal} {\bibinfo
  {journal} {JCAP}\ }\textbf {\bibinfo {volume} {05}},\ \bibinfo {pages} {021}
  (\bibinfo {year} {2023})},\ \Eprint {http://arxiv.org/abs/2209.05926}
  {arXiv:2209.05926 [astro-ph.HE]} \BibitemShut {NoStop}%
\bibitem [{\citenamefont {Cao}\ \emph {et~al.}(2023)\citenamefont {Cao} \emph
  {et~al.}}]{LHAASO:2023gne}%
  \BibitemOpen
  \bibfield  {author} {\bibinfo {author} {\bibfnamefont {Z.}~\bibnamefont
  {Cao}} \emph {et~al.} (\bibinfo {collaboration} {LHAASO}),\ }\href {\doibase
  10.1103/PhysRevLett.131.151001} {\bibfield  {journal} {\bibinfo  {journal}
  {Phys. Rev. Lett.}\ }\textbf {\bibinfo {volume} {131}},\ \bibinfo {pages}
  {151001} (\bibinfo {year} {2023})},\ \Eprint
  {http://arxiv.org/abs/2305.05372} {arXiv:2305.05372 [astro-ph.HE]}
  \BibitemShut {NoStop}%
\bibitem [{\citenamefont {Cao}\ \emph {et~al.}(2022)\citenamefont {Cao} \emph
  {et~al.}}]{LHAASO:2022yxw}%
  \BibitemOpen
  \bibfield  {author} {\bibinfo {author} {\bibfnamefont {Z.}~\bibnamefont
  {Cao}} \emph {et~al.} (\bibinfo {collaboration} {LHAASO}),\ }\href {\doibase
  10.1103/PhysRevLett.129.261103} {\bibfield  {journal} {\bibinfo  {journal}
  {Phys. Rev. Lett.}\ }\textbf {\bibinfo {volume} {129}},\ \bibinfo {pages}
  {261103} (\bibinfo {year} {2022})},\ \Eprint
  {http://arxiv.org/abs/2210.15989} {arXiv:2210.15989 [astro-ph.HE]}
  \BibitemShut {NoStop}%
\bibitem [{\citenamefont {Das}\ \emph {et~al.}(2023)\citenamefont {Das},
  \citenamefont {Murase},\ and\ \citenamefont {Fujii}}]{Das:2023wtk}%
  \BibitemOpen
  \bibfield  {author} {\bibinfo {author} {\bibfnamefont {S.}~\bibnamefont
  {Das}}, \bibinfo {author} {\bibfnamefont {K.}~\bibnamefont {Murase}}, \ and\
  \bibinfo {author} {\bibfnamefont {T.}~\bibnamefont {Fujii}},\ }\href
  {\doibase 10.1103/PhysRevD.107.103013} {\bibfield  {journal} {\bibinfo
  {journal} {Phys. Rev. D}\ }\textbf {\bibinfo {volume} {107}},\ \bibinfo
  {pages} {103013} (\bibinfo {year} {2023})},\ \Eprint
  {http://arxiv.org/abs/2302.02993} {arXiv:2302.02993 [astro-ph.HE]}
  \BibitemShut {NoStop}%
\bibitem [{\citenamefont {Adriani}\ \emph
  {et~al.}(2025{\natexlab{c}})\citenamefont {Adriani} \emph
  {et~al.}}]{KM3NeT:2025ccp}%
  \BibitemOpen
  \bibfield  {author} {\bibinfo {author} {\bibfnamefont {O.}~\bibnamefont
  {Adriani}} \emph {et~al.} (\bibinfo {collaboration} {KM3NeT}),\ }\href@noop
  {} {\  (\bibinfo {year} {2025}{\natexlab{c}})},\ \Eprint
  {http://arxiv.org/abs/2502.08173} {arXiv:2502.08173 [astro-ph.HE]}
  \BibitemShut {NoStop}%
\bibitem [{\citenamefont {Aartsen}\ \emph {et~al.}(2018)\citenamefont {Aartsen}
  \emph {et~al.}}]{IceCube:2018fhm}%
  \BibitemOpen
  \bibfield  {author} {\bibinfo {author} {\bibfnamefont {M.~G.}\ \bibnamefont
  {Aartsen}} \emph {et~al.} (\bibinfo {collaboration} {IceCube}),\ }\href
  {\doibase 10.1103/PhysRevD.98.062003} {\bibfield  {journal} {\bibinfo
  {journal} {Phys. Rev. D}\ }\textbf {\bibinfo {volume} {98}},\ \bibinfo
  {pages} {062003} (\bibinfo {year} {2018})},\ \Eprint
  {http://arxiv.org/abs/1807.01820} {arXiv:1807.01820 [astro-ph.HE]}
  \BibitemShut {NoStop}%
\bibitem [{\citenamefont {Abbasi}\ \emph {et~al.}(2021)\citenamefont {Abbasi}
  \emph {et~al.}}]{IceCube:2020wum}%
  \BibitemOpen
  \bibfield  {author} {\bibinfo {author} {\bibfnamefont {R.}~\bibnamefont
  {Abbasi}} \emph {et~al.} (\bibinfo {collaboration} {IceCube}),\ }\href
  {\doibase 10.1103/PhysRevD.104.022002} {\bibfield  {journal} {\bibinfo
  {journal} {Phys. Rev. D}\ }\textbf {\bibinfo {volume} {104}},\ \bibinfo
  {pages} {022002} (\bibinfo {year} {2021})},\ \Eprint
  {http://arxiv.org/abs/2011.03545} {arXiv:2011.03545 [astro-ph.HE]}
  \BibitemShut {NoStop}%
\bibitem [{\citenamefont {Abdul~Halim}(2023)}]{AbdulHalim:2023SN}%
  \BibitemOpen
  \bibfield  {author} {\bibinfo {author} {\bibfnamefont {e.~a.}\ \bibnamefont
  {Abdul~Halim}},\ }\href {\doibase 10.22323/1.444.1488} {\bibfield  {journal}
  {\bibinfo  {journal} {PoS}\ }\textbf {\bibinfo {volume} {ICRC2023}},\
  \bibinfo {pages} {1488} (\bibinfo {year} {2023})}\BibitemShut {NoStop}%
\bibitem [{\citenamefont {Aiello}\ \emph {et~al.}(2019)\citenamefont {Aiello}
  \emph {et~al.}}]{KM3NeT:2018wnd}%
  \BibitemOpen
  \bibfield  {author} {\bibinfo {author} {\bibfnamefont {S.}~\bibnamefont
  {Aiello}} \emph {et~al.} (\bibinfo {collaboration} {KM3NeT}),\ }\href
  {\doibase 10.1016/j.astropartphys.2019.04.002} {\bibfield  {journal}
  {\bibinfo  {journal} {Astropart. Phys.}\ }\textbf {\bibinfo {volume} {111}},\
  \bibinfo {pages} {100} (\bibinfo {year} {2019})},\ \Eprint
  {http://arxiv.org/abs/1810.08499} {arXiv:1810.08499 [astro-ph.HE]}
  \BibitemShut {NoStop}%
\bibitem [{\citenamefont {Li}\ \emph {et~al.}(2025)\citenamefont {Li},
  \citenamefont {Machado}, \citenamefont {Naredo-Tuero},\ and\ \citenamefont
  {Schwemberger}}]{Li:2025tqf}%
  \BibitemOpen
  \bibfield  {author} {\bibinfo {author} {\bibfnamefont {S.~W.}\ \bibnamefont
  {Li}}, \bibinfo {author} {\bibfnamefont {P.}~\bibnamefont {Machado}},
  \bibinfo {author} {\bibfnamefont {D.}~\bibnamefont {Naredo-Tuero}}, \ and\
  \bibinfo {author} {\bibfnamefont {T.}~\bibnamefont {Schwemberger}},\
  }\href@noop {} {\  (\bibinfo {year} {2025})},\ \Eprint
  {http://arxiv.org/abs/2502.04508} {arXiv:2502.04508 [astro-ph.HE]}
  \BibitemShut {NoStop}%
\bibitem [{\citenamefont {Brdar}\ and\ \citenamefont
  {Chattopadhyay}(2025)}]{Brdar:2025azm}%
  \BibitemOpen
  \bibfield  {author} {\bibinfo {author} {\bibfnamefont {V.}~\bibnamefont
  {Brdar}}\ and\ \bibinfo {author} {\bibfnamefont {D.~S.}\ \bibnamefont
  {Chattopadhyay}},\ }\href@noop {} {\  (\bibinfo {year} {2025})},\ \Eprint
  {http://arxiv.org/abs/2502.21299} {arXiv:2502.21299 [hep-ph]} \BibitemShut
  {NoStop}%
\bibitem [{\citenamefont {Abazajian}\ \emph {et~al.}(2012)\citenamefont
  {Abazajian} \emph {et~al.}}]{Abazajian:2012ys}%
  \BibitemOpen
  \bibfield  {author} {\bibinfo {author} {\bibfnamefont {K.~N.}\ \bibnamefont
  {Abazajian}} \emph {et~al.},\ }\href@noop {} {\  (\bibinfo {year} {2012})},\
  \Eprint {http://arxiv.org/abs/1204.5379} {arXiv:1204.5379 [hep-ph]}
  \BibitemShut {NoStop}%
\bibitem [{\citenamefont {Boyarsky}\ \emph {et~al.}(2009)\citenamefont
  {Boyarsky}, \citenamefont {Ruchayskiy},\ and\ \citenamefont
  {Shaposhnikov}}]{Boyarsky:2009ix}%
  \BibitemOpen
  \bibfield  {author} {\bibinfo {author} {\bibfnamefont {A.}~\bibnamefont
  {Boyarsky}}, \bibinfo {author} {\bibfnamefont {O.}~\bibnamefont
  {Ruchayskiy}}, \ and\ \bibinfo {author} {\bibfnamefont {M.}~\bibnamefont
  {Shaposhnikov}},\ }\href {\doibase 10.1146/annurev.nucl.010909.083654}
  {\bibfield  {journal} {\bibinfo  {journal} {Ann. Rev. Nucl. Part. Sci.}\
  }\textbf {\bibinfo {volume} {59}},\ \bibinfo {pages} {191} (\bibinfo {year}
  {2009})},\ \Eprint {http://arxiv.org/abs/0901.0011} {arXiv:0901.0011
  [hep-ph]} \BibitemShut {NoStop}%
\bibitem [{\citenamefont {Drewes}\ \emph {et~al.}(2017)\citenamefont {Drewes}
  \emph {et~al.}}]{Drewes:2016upu}%
  \BibitemOpen
  \bibfield  {author} {\bibinfo {author} {\bibfnamefont {M.}~\bibnamefont
  {Drewes}} \emph {et~al.},\ }\href {\doibase 10.1088/1475-7516/2017/01/025}
  {\bibfield  {journal} {\bibinfo  {journal} {JCAP}\ }\textbf {\bibinfo
  {volume} {01}},\ \bibinfo {pages} {025} (\bibinfo {year} {2017})},\ \Eprint
  {http://arxiv.org/abs/1602.04816} {arXiv:1602.04816 [hep-ph]} \BibitemShut
  {NoStop}%
\bibitem [{\citenamefont {Abazajian}(2017)}]{Abazajian:2017tcc}%
  \BibitemOpen
  \bibfield  {author} {\bibinfo {author} {\bibfnamefont {K.~N.}\ \bibnamefont
  {Abazajian}},\ }\href {\doibase 10.1016/j.physrep.2017.10.003} {\bibfield
  {journal} {\bibinfo  {journal} {Phys. Rept.}\ }\textbf {\bibinfo {volume}
  {711-712}},\ \bibinfo {pages} {1} (\bibinfo {year} {2017})},\ \Eprint
  {http://arxiv.org/abs/1705.01837} {arXiv:1705.01837 [hep-ph]} \BibitemShut
  {NoStop}%
\bibitem [{\citenamefont {Acero}\ \emph {et~al.}(2024)\citenamefont {Acero}
  \emph {et~al.}}]{Acero:2022wqg}%
  \BibitemOpen
  \bibfield  {author} {\bibinfo {author} {\bibfnamefont {M.~A.}\ \bibnamefont
  {Acero}} \emph {et~al.},\ }\href {\doibase 10.1088/1361-6471/ad307f}
  {\bibfield  {journal} {\bibinfo  {journal} {J. Phys. G}\ }\textbf {\bibinfo
  {volume} {51}},\ \bibinfo {pages} {120501} (\bibinfo {year} {2024})},\
  \Eprint {http://arxiv.org/abs/2203.07323} {arXiv:2203.07323 [hep-ex]}
  \BibitemShut {NoStop}%
\bibitem [{\citenamefont {Cherry}\ and\ \citenamefont
  {Shoemaker}(2019)}]{Cherry:2018rxj}%
  \BibitemOpen
  \bibfield  {author} {\bibinfo {author} {\bibfnamefont {J.~F.}\ \bibnamefont
  {Cherry}}\ and\ \bibinfo {author} {\bibfnamefont {I.~M.}\ \bibnamefont
  {Shoemaker}},\ }\href {\doibase 10.1103/PhysRevD.99.063016} {\bibfield
  {journal} {\bibinfo  {journal} {Phys. Rev. D}\ }\textbf {\bibinfo {volume}
  {99}},\ \bibinfo {pages} {063016} (\bibinfo {year} {2019})},\ \Eprint
  {http://arxiv.org/abs/1802.01611} {arXiv:1802.01611 [hep-ph]} \BibitemShut
  {NoStop}%
\bibitem [{\citenamefont {Huang}(2018)}]{Huang:2018als}%
  \BibitemOpen
  \bibfield  {author} {\bibinfo {author} {\bibfnamefont {G.-y.}\ \bibnamefont
  {Huang}},\ }\href {\doibase 10.1103/PhysRevD.98.043019} {\bibfield  {journal}
  {\bibinfo  {journal} {Phys. Rev. D}\ }\textbf {\bibinfo {volume} {98}},\
  \bibinfo {pages} {043019} (\bibinfo {year} {2018})},\ \Eprint
  {http://arxiv.org/abs/1804.05362} {arXiv:1804.05362 [hep-ph]} \BibitemShut
  {NoStop}%
\bibitem [{\citenamefont {Jaeckel}\ and\ \citenamefont
  {Yin}(2021{\natexlab{a}})}]{Jaeckel:2020oet}%
  \BibitemOpen
  \bibfield  {author} {\bibinfo {author} {\bibfnamefont {J.}~\bibnamefont
  {Jaeckel}}\ and\ \bibinfo {author} {\bibfnamefont {W.}~\bibnamefont {Yin}},\
  }\href {\doibase 10.1088/1475-7516/2021/02/044} {\bibfield  {journal}
  {\bibinfo  {journal} {JCAP}\ }\textbf {\bibinfo {volume} {02}},\ \bibinfo
  {pages} {044} (\bibinfo {year} {2021}{\natexlab{a}})},\ \Eprint
  {http://arxiv.org/abs/2007.15006} {arXiv:2007.15006 [hep-ph]} \BibitemShut
  {NoStop}%
\bibitem [{\citenamefont {Jaeckel}\ and\ \citenamefont
  {Yin}(2022)}]{Jaeckel:2021ert}%
  \BibitemOpen
  \bibfield  {author} {\bibinfo {author} {\bibfnamefont {J.}~\bibnamefont
  {Jaeckel}}\ and\ \bibinfo {author} {\bibfnamefont {W.}~\bibnamefont {Yin}},\
  }\href {\doibase 10.1103/PhysRevD.105.115003} {\bibfield  {journal} {\bibinfo
   {journal} {Phys. Rev. D}\ }\textbf {\bibinfo {volume} {105}},\ \bibinfo
  {pages} {115003} (\bibinfo {year} {2022})},\ \Eprint
  {http://arxiv.org/abs/2110.03692} {arXiv:2110.03692 [hep-ph]} \BibitemShut
  {NoStop}%
\bibitem [{\citenamefont {Minkowski}(1977)}]{Minkowski:1977sc}%
  \BibitemOpen
  \bibfield  {author} {\bibinfo {author} {\bibfnamefont {P.}~\bibnamefont
  {Minkowski}},\ }\href {\doibase 10.1016/0370-2693(77)90435-X} {\bibfield
  {journal} {\bibinfo  {journal} {Phys. Lett. B}\ }\textbf {\bibinfo {volume}
  {67}},\ \bibinfo {pages} {421} (\bibinfo {year} {1977})}\BibitemShut
  {NoStop}%
\bibitem [{\citenamefont {Yanagida}(1979)}]{Yanagida:1979as}%
  \BibitemOpen
  \bibfield  {author} {\bibinfo {author} {\bibfnamefont {T.}~\bibnamefont
  {Yanagida}},\ }\href@noop {} {\bibfield  {journal} {\bibinfo  {journal}
  {Conf. Proc. C}\ }\textbf {\bibinfo {volume} {7902131}},\ \bibinfo {pages}
  {95} (\bibinfo {year} {1979})}\BibitemShut {NoStop}%
\bibitem [{\citenamefont {Glashow}(1980)}]{Glashow:1979nm}%
  \BibitemOpen
  \bibfield  {author} {\bibinfo {author} {\bibfnamefont {S.~L.}\ \bibnamefont
  {Glashow}},\ }\href {\doibase 10.1007/978-1-4684-7197-7_15} {\bibfield
  {journal} {\bibinfo  {journal} {NATO Sci. Ser. B}\ }\textbf {\bibinfo
  {volume} {61}},\ \bibinfo {pages} {687} (\bibinfo {year} {1980})}\BibitemShut
  {NoStop}%
\bibitem [{\citenamefont {Gell-Mann}\ \emph {et~al.}(1979)\citenamefont
  {Gell-Mann}, \citenamefont {Ramond},\ and\ \citenamefont
  {Slansky}}]{GellMann:1980vs}%
  \BibitemOpen
  \bibfield  {author} {\bibinfo {author} {\bibfnamefont {M.}~\bibnamefont
  {Gell-Mann}}, \bibinfo {author} {\bibfnamefont {P.}~\bibnamefont {Ramond}}, \
  and\ \bibinfo {author} {\bibfnamefont {R.}~\bibnamefont {Slansky}},\
  }\href@noop {} {\bibfield  {journal} {\bibinfo  {journal} {Conf. Proc. C}\
  }\textbf {\bibinfo {volume} {790927}},\ \bibinfo {pages} {315} (\bibinfo
  {year} {1979})},\ \Eprint {http://arxiv.org/abs/1306.4669} {arXiv:1306.4669
  [hep-th]} \BibitemShut {NoStop}%
\bibitem [{\citenamefont {Mohapatra}\ and\ \citenamefont
  {Senjanovic}(1980)}]{Mohapatra:1979ia}%
  \BibitemOpen
  \bibfield  {author} {\bibinfo {author} {\bibfnamefont {R.~N.}\ \bibnamefont
  {Mohapatra}}\ and\ \bibinfo {author} {\bibfnamefont {G.}~\bibnamefont
  {Senjanovic}},\ }\href {\doibase 10.1103/PhysRevLett.44.912} {\bibfield
  {journal} {\bibinfo  {journal} {Phys. Rev. Lett.}\ }\textbf {\bibinfo
  {volume} {44}},\ \bibinfo {pages} {912} (\bibinfo {year} {1980})}\BibitemShut
  {NoStop}%
\bibitem [{\citenamefont {Langhoff}\ \emph {et~al.}(2022)\citenamefont
  {Langhoff}, \citenamefont {Outmezguine},\ and\ \citenamefont
  {Rodd}}]{Langhoff:2022bij}%
  \BibitemOpen
  \bibfield  {author} {\bibinfo {author} {\bibfnamefont {K.}~\bibnamefont
  {Langhoff}}, \bibinfo {author} {\bibfnamefont {N.~J.}\ \bibnamefont
  {Outmezguine}}, \ and\ \bibinfo {author} {\bibfnamefont {N.~L.}\ \bibnamefont
  {Rodd}},\ }\href {\doibase 10.1103/PhysRevLett.129.241101} {\bibfield
  {journal} {\bibinfo  {journal} {Phys. Rev. Lett.}\ }\textbf {\bibinfo
  {volume} {129}},\ \bibinfo {pages} {241101} (\bibinfo {year} {2022})},\
  \Eprint {http://arxiv.org/abs/2209.06216} {arXiv:2209.06216 [hep-ph]}
  \BibitemShut {NoStop}%
\bibitem [{\citenamefont {Cadamuro}\ and\ \citenamefont
  {Redondo}(2012)}]{Cadamuro:2011fd}%
  \BibitemOpen
  \bibfield  {author} {\bibinfo {author} {\bibfnamefont {D.}~\bibnamefont
  {Cadamuro}}\ and\ \bibinfo {author} {\bibfnamefont {J.}~\bibnamefont
  {Redondo}},\ }\href {\doibase 10.1088/1475-7516/2012/02/032} {\bibfield
  {journal} {\bibinfo  {journal} {JCAP}\ }\textbf {\bibinfo {volume} {02}},\
  \bibinfo {pages} {032} (\bibinfo {year} {2012})},\ \Eprint
  {http://arxiv.org/abs/1110.2895} {arXiv:1110.2895 [hep-ph]} \BibitemShut
  {NoStop}%
\bibitem [{\citenamefont {Pal}\ and\ \citenamefont
  {Wolfenstein}(1982)}]{Pal:1981rm}%
  \BibitemOpen
  \bibfield  {author} {\bibinfo {author} {\bibfnamefont {P.~B.}\ \bibnamefont
  {Pal}}\ and\ \bibinfo {author} {\bibfnamefont {L.}~\bibnamefont
  {Wolfenstein}},\ }\href {\doibase 10.1103/PhysRevD.25.766} {\bibfield
  {journal} {\bibinfo  {journal} {Phys. Rev. D}\ }\textbf {\bibinfo {volume}
  {25}},\ \bibinfo {pages} {766} (\bibinfo {year} {1982})}\BibitemShut
  {NoStop}%
\bibitem [{\citenamefont {Boyarsky}\ \emph {et~al.}(2019)\citenamefont
  {Boyarsky}, \citenamefont {Drewes}, \citenamefont {Lasserre}, \citenamefont
  {Mertens},\ and\ \citenamefont {Ruchayskiy}}]{Boyarsky:2018tvu}%
  \BibitemOpen
  \bibfield  {author} {\bibinfo {author} {\bibfnamefont {A.}~\bibnamefont
  {Boyarsky}}, \bibinfo {author} {\bibfnamefont {M.}~\bibnamefont {Drewes}},
  \bibinfo {author} {\bibfnamefont {T.}~\bibnamefont {Lasserre}}, \bibinfo
  {author} {\bibfnamefont {S.}~\bibnamefont {Mertens}}, \ and\ \bibinfo
  {author} {\bibfnamefont {O.}~\bibnamefont {Ruchayskiy}},\ }\href {\doibase
  10.1016/j.ppnp.2018.07.004} {\bibfield  {journal} {\bibinfo  {journal} {Prog.
  Part. Nucl. Phys.}\ }\textbf {\bibinfo {volume} {104}},\ \bibinfo {pages} {1}
  (\bibinfo {year} {2019})},\ \Eprint {http://arxiv.org/abs/1807.07938}
  {arXiv:1807.07938 [hep-ph]} \BibitemShut {NoStop}%
\bibitem [{\citenamefont {Gelmini}\ \emph {et~al.}(2004)\citenamefont
  {Gelmini}, \citenamefont {Palomares-Ruiz},\ and\ \citenamefont
  {Pascoli}}]{Gelmini:2004ah}%
  \BibitemOpen
  \bibfield  {author} {\bibinfo {author} {\bibfnamefont {G.}~\bibnamefont
  {Gelmini}}, \bibinfo {author} {\bibfnamefont {S.}~\bibnamefont
  {Palomares-Ruiz}}, \ and\ \bibinfo {author} {\bibfnamefont {S.}~\bibnamefont
  {Pascoli}},\ }\href {\doibase 10.1103/PhysRevLett.93.081302} {\bibfield
  {journal} {\bibinfo  {journal} {Phys. Rev. Lett.}\ }\textbf {\bibinfo
  {volume} {93}},\ \bibinfo {pages} {081302} (\bibinfo {year} {2004})},\
  \Eprint {http://arxiv.org/abs/astro-ph/0403323} {arXiv:astro-ph/0403323}
  \BibitemShut {NoStop}%
\bibitem [{\citenamefont {Dodelson}\ and\ \citenamefont
  {Widrow}(1994)}]{Dodelson:1993je}%
  \BibitemOpen
  \bibfield  {author} {\bibinfo {author} {\bibfnamefont {S.}~\bibnamefont
  {Dodelson}}\ and\ \bibinfo {author} {\bibfnamefont {L.~M.}\ \bibnamefont
  {Widrow}},\ }\href {\doibase 10.1103/PhysRevLett.72.17} {\bibfield  {journal}
  {\bibinfo  {journal} {Phys. Rev. Lett.}\ }\textbf {\bibinfo {volume} {72}},\
  \bibinfo {pages} {17} (\bibinfo {year} {1994})},\ \Eprint
  {http://arxiv.org/abs/hep-ph/9303287} {arXiv:hep-ph/9303287} \BibitemShut
  {NoStop}%
\bibitem [{\citenamefont {Hasegawa}\ \emph {et~al.}(2019)\citenamefont
  {Hasegawa}, \citenamefont {Hiroshima}, \citenamefont {Kohri}, \citenamefont
  {Hansen}, \citenamefont {Tram},\ and\ \citenamefont
  {Hannestad}}]{Hasegawa:2019jsa}%
  \BibitemOpen
  \bibfield  {author} {\bibinfo {author} {\bibfnamefont {T.}~\bibnamefont
  {Hasegawa}}, \bibinfo {author} {\bibfnamefont {N.}~\bibnamefont {Hiroshima}},
  \bibinfo {author} {\bibfnamefont {K.}~\bibnamefont {Kohri}}, \bibinfo
  {author} {\bibfnamefont {R.~S.~L.}\ \bibnamefont {Hansen}}, \bibinfo {author}
  {\bibfnamefont {T.}~\bibnamefont {Tram}}, \ and\ \bibinfo {author}
  {\bibfnamefont {S.}~\bibnamefont {Hannestad}},\ }\href {\doibase
  10.1088/1475-7516/2019/12/012} {\bibfield  {journal} {\bibinfo  {journal}
  {JCAP}\ }\textbf {\bibinfo {volume} {12}},\ \bibinfo {pages} {012} (\bibinfo
  {year} {2019})},\ \Eprint {http://arxiv.org/abs/1908.10189} {arXiv:1908.10189
  [hep-ph]} \BibitemShut {NoStop}%
\bibitem [{\citenamefont {Kawasaki}\ \emph {et~al.}(1999)\citenamefont
  {Kawasaki}, \citenamefont {Kohri},\ and\ \citenamefont
  {Sugiyama}}]{Kawasaki:1999na}%
  \BibitemOpen
  \bibfield  {author} {\bibinfo {author} {\bibfnamefont {M.}~\bibnamefont
  {Kawasaki}}, \bibinfo {author} {\bibfnamefont {K.}~\bibnamefont {Kohri}}, \
  and\ \bibinfo {author} {\bibfnamefont {N.}~\bibnamefont {Sugiyama}},\ }\href
  {\doibase 10.1103/PhysRevLett.82.4168} {\bibfield  {journal} {\bibinfo
  {journal} {Phys. Rev. Lett.}\ }\textbf {\bibinfo {volume} {82}},\ \bibinfo
  {pages} {4168} (\bibinfo {year} {1999})},\ \Eprint
  {http://arxiv.org/abs/astro-ph/9811437} {arXiv:astro-ph/9811437} \BibitemShut
  {NoStop}%
\bibitem [{\citenamefont {Kawasaki}\ \emph {et~al.}(2000)\citenamefont
  {Kawasaki}, \citenamefont {Kohri},\ and\ \citenamefont
  {Sugiyama}}]{Kawasaki:2000en}%
  \BibitemOpen
  \bibfield  {author} {\bibinfo {author} {\bibfnamefont {M.}~\bibnamefont
  {Kawasaki}}, \bibinfo {author} {\bibfnamefont {K.}~\bibnamefont {Kohri}}, \
  and\ \bibinfo {author} {\bibfnamefont {N.}~\bibnamefont {Sugiyama}},\ }\href
  {\doibase 10.1103/PhysRevD.62.023506} {\bibfield  {journal} {\bibinfo
  {journal} {Phys. Rev. D}\ }\textbf {\bibinfo {volume} {62}},\ \bibinfo
  {pages} {023506} (\bibinfo {year} {2000})},\ \Eprint
  {http://arxiv.org/abs/astro-ph/0002127} {arXiv:astro-ph/0002127} \BibitemShut
  {NoStop}%
\bibitem [{\citenamefont {Hannestad}(2004)}]{Hannestad:2004px}%
  \BibitemOpen
  \bibfield  {author} {\bibinfo {author} {\bibfnamefont {S.}~\bibnamefont
  {Hannestad}},\ }\href {\doibase 10.1103/PhysRevD.70.043506} {\bibfield
  {journal} {\bibinfo  {journal} {Phys. Rev. D}\ }\textbf {\bibinfo {volume}
  {70}},\ \bibinfo {pages} {043506} (\bibinfo {year} {2004})},\ \Eprint
  {http://arxiv.org/abs/astro-ph/0403291} {arXiv:astro-ph/0403291} \BibitemShut
  {NoStop}%
\bibitem [{\citenamefont {Ichikawa}\ \emph {et~al.}(2007)\citenamefont
  {Ichikawa}, \citenamefont {Kawasaki},\ and\ \citenamefont
  {Takahashi}}]{Ichikawa:2006vm}%
  \BibitemOpen
  \bibfield  {author} {\bibinfo {author} {\bibfnamefont {K.}~\bibnamefont
  {Ichikawa}}, \bibinfo {author} {\bibfnamefont {M.}~\bibnamefont {Kawasaki}},
  \ and\ \bibinfo {author} {\bibfnamefont {F.}~\bibnamefont {Takahashi}},\
  }\href {\doibase 10.1088/1475-7516/2007/05/007} {\bibfield  {journal}
  {\bibinfo  {journal} {JCAP}\ }\textbf {\bibinfo {volume} {05}},\ \bibinfo
  {pages} {007} (\bibinfo {year} {2007})},\ \Eprint
  {http://arxiv.org/abs/astro-ph/0611784} {arXiv:astro-ph/0611784} \BibitemShut
  {NoStop}%
\bibitem [{\citenamefont {De~Bernardis}\ \emph {et~al.}(2008)\citenamefont
  {De~Bernardis}, \citenamefont {Pagano},\ and\ \citenamefont
  {Melchiorri}}]{DeBernardis:2008zz}%
  \BibitemOpen
  \bibfield  {author} {\bibinfo {author} {\bibfnamefont {F.}~\bibnamefont
  {De~Bernardis}}, \bibinfo {author} {\bibfnamefont {L.}~\bibnamefont
  {Pagano}}, \ and\ \bibinfo {author} {\bibfnamefont {A.}~\bibnamefont
  {Melchiorri}},\ }\href {\doibase 10.1016/j.astropartphys.2008.09.005}
  {\bibfield  {journal} {\bibinfo  {journal} {Astropart. Phys.}\ }\textbf
  {\bibinfo {volume} {30}},\ \bibinfo {pages} {192} (\bibinfo {year}
  {2008})}\BibitemShut {NoStop}%
\bibitem [{\citenamefont {de~Salas}\ \emph {et~al.}(2015)\citenamefont
  {de~Salas}, \citenamefont {Lattanzi}, \citenamefont {Mangano}, \citenamefont
  {Miele}, \citenamefont {Pastor},\ and\ \citenamefont
  {Pisanti}}]{deSalas:2015glj}%
  \BibitemOpen
  \bibfield  {author} {\bibinfo {author} {\bibfnamefont {P.~F.}\ \bibnamefont
  {de~Salas}}, \bibinfo {author} {\bibfnamefont {M.}~\bibnamefont {Lattanzi}},
  \bibinfo {author} {\bibfnamefont {G.}~\bibnamefont {Mangano}}, \bibinfo
  {author} {\bibfnamefont {G.}~\bibnamefont {Miele}}, \bibinfo {author}
  {\bibfnamefont {S.}~\bibnamefont {Pastor}}, \ and\ \bibinfo {author}
  {\bibfnamefont {O.}~\bibnamefont {Pisanti}},\ }\href {\doibase
  10.1103/PhysRevD.92.123534} {\bibfield  {journal} {\bibinfo  {journal} {Phys.
  Rev. D}\ }\textbf {\bibinfo {volume} {92}},\ \bibinfo {pages} {123534}
  (\bibinfo {year} {2015})},\ \Eprint {http://arxiv.org/abs/1511.00672}
  {arXiv:1511.00672 [astro-ph.CO]} \BibitemShut {NoStop}%
\bibitem [{\citenamefont {Gewering-Peine}\ \emph {et~al.}(2017)\citenamefont
  {Gewering-Peine}, \citenamefont {Horns},\ and\ \citenamefont
  {Schmitt}}]{Gewering-Peine:2016yoj}%
  \BibitemOpen
  \bibfield  {author} {\bibinfo {author} {\bibfnamefont {A.}~\bibnamefont
  {Gewering-Peine}}, \bibinfo {author} {\bibfnamefont {D.}~\bibnamefont
  {Horns}}, \ and\ \bibinfo {author} {\bibfnamefont {J.~H. M.~M.}\ \bibnamefont
  {Schmitt}},\ }\href {\doibase 10.1088/1475-7516/2017/06/036} {\bibfield
  {journal} {\bibinfo  {journal} {JCAP}\ }\textbf {\bibinfo {volume} {06}},\
  \bibinfo {pages} {036} (\bibinfo {year} {2017})},\ \Eprint
  {http://arxiv.org/abs/1611.01733} {arXiv:1611.01733 [astro-ph.HE]}
  \BibitemShut {NoStop}%
\bibitem [{\citenamefont {Foster}\ \emph {et~al.}(2021)\citenamefont {Foster},
  \citenamefont {Kongsore}, \citenamefont {Dessert}, \citenamefont {Park},
  \citenamefont {Rodd}, \citenamefont {Cranmer},\ and\ \citenamefont
  {Safdi}}]{Foster:2021ngm}%
  \BibitemOpen
  \bibfield  {author} {\bibinfo {author} {\bibfnamefont {J.~W.}\ \bibnamefont
  {Foster}}, \bibinfo {author} {\bibfnamefont {M.}~\bibnamefont {Kongsore}},
  \bibinfo {author} {\bibfnamefont {C.}~\bibnamefont {Dessert}}, \bibinfo
  {author} {\bibfnamefont {Y.}~\bibnamefont {Park}}, \bibinfo {author}
  {\bibfnamefont {N.~L.}\ \bibnamefont {Rodd}}, \bibinfo {author}
  {\bibfnamefont {K.}~\bibnamefont {Cranmer}}, \ and\ \bibinfo {author}
  {\bibfnamefont {B.~R.}\ \bibnamefont {Safdi}},\ }\href {\doibase
  10.1103/PhysRevLett.127.051101} {\bibfield  {journal} {\bibinfo  {journal}
  {Phys. Rev. Lett.}\ }\textbf {\bibinfo {volume} {127}},\ \bibinfo {pages}
  {051101} (\bibinfo {year} {2021})},\ \Eprint
  {http://arxiv.org/abs/2102.02207} {arXiv:2102.02207 [astro-ph.CO]}
  \BibitemShut {NoStop}%
\bibitem [{\citenamefont {Perez}\ \emph {et~al.}(2017)\citenamefont {Perez},
  \citenamefont {Ng}, \citenamefont {Beacom}, \citenamefont {Hersh},
  \citenamefont {Horiuchi},\ and\ \citenamefont {Krivonos}}]{Perez:2016tcq}%
  \BibitemOpen
  \bibfield  {author} {\bibinfo {author} {\bibfnamefont {K.}~\bibnamefont
  {Perez}}, \bibinfo {author} {\bibfnamefont {K.~C.~Y.}\ \bibnamefont {Ng}},
  \bibinfo {author} {\bibfnamefont {J.~F.}\ \bibnamefont {Beacom}}, \bibinfo
  {author} {\bibfnamefont {C.}~\bibnamefont {Hersh}}, \bibinfo {author}
  {\bibfnamefont {S.}~\bibnamefont {Horiuchi}}, \ and\ \bibinfo {author}
  {\bibfnamefont {R.}~\bibnamefont {Krivonos}},\ }\href {\doibase
  10.1103/PhysRevD.95.123002} {\bibfield  {journal} {\bibinfo  {journal} {Phys.
  Rev. D}\ }\textbf {\bibinfo {volume} {95}},\ \bibinfo {pages} {123002}
  (\bibinfo {year} {2017})},\ \Eprint {http://arxiv.org/abs/1609.00667}
  {arXiv:1609.00667 [astro-ph.HE]} \BibitemShut {NoStop}%
\bibitem [{\citenamefont {Ng}\ \emph {et~al.}(2019)\citenamefont {Ng},
  \citenamefont {Roach}, \citenamefont {Perez}, \citenamefont {Beacom},
  \citenamefont {Horiuchi}, \citenamefont {Krivonos},\ and\ \citenamefont
  {Wik}}]{Ng:2019gch}%
  \BibitemOpen
  \bibfield  {author} {\bibinfo {author} {\bibfnamefont {K.~C.~Y.}\
  \bibnamefont {Ng}}, \bibinfo {author} {\bibfnamefont {B.~M.}\ \bibnamefont
  {Roach}}, \bibinfo {author} {\bibfnamefont {K.}~\bibnamefont {Perez}},
  \bibinfo {author} {\bibfnamefont {J.~F.}\ \bibnamefont {Beacom}}, \bibinfo
  {author} {\bibfnamefont {S.}~\bibnamefont {Horiuchi}}, \bibinfo {author}
  {\bibfnamefont {R.}~\bibnamefont {Krivonos}}, \ and\ \bibinfo {author}
  {\bibfnamefont {D.~R.}\ \bibnamefont {Wik}},\ }\href {\doibase
  10.1103/PhysRevD.99.083005} {\bibfield  {journal} {\bibinfo  {journal} {Phys.
  Rev. D}\ }\textbf {\bibinfo {volume} {99}},\ \bibinfo {pages} {083005}
  (\bibinfo {year} {2019})},\ \Eprint {http://arxiv.org/abs/1901.01262}
  {arXiv:1901.01262 [astro-ph.HE]} \BibitemShut {NoStop}%
\bibitem [{\citenamefont {Roach}\ \emph {et~al.}(2023)\citenamefont {Roach},
  \citenamefont {Rossland}, \citenamefont {Ng}, \citenamefont {Perez},
  \citenamefont {Beacom}, \citenamefont {Grefenstette}, \citenamefont
  {Horiuchi}, \citenamefont {Krivonos},\ and\ \citenamefont
  {Wik}}]{Roach:2022lgo}%
  \BibitemOpen
  \bibfield  {author} {\bibinfo {author} {\bibfnamefont {B.~M.}\ \bibnamefont
  {Roach}}, \bibinfo {author} {\bibfnamefont {S.}~\bibnamefont {Rossland}},
  \bibinfo {author} {\bibfnamefont {K.~C.~Y.}\ \bibnamefont {Ng}}, \bibinfo
  {author} {\bibfnamefont {K.}~\bibnamefont {Perez}}, \bibinfo {author}
  {\bibfnamefont {J.~F.}\ \bibnamefont {Beacom}}, \bibinfo {author}
  {\bibfnamefont {B.~W.}\ \bibnamefont {Grefenstette}}, \bibinfo {author}
  {\bibfnamefont {S.}~\bibnamefont {Horiuchi}}, \bibinfo {author}
  {\bibfnamefont {R.}~\bibnamefont {Krivonos}}, \ and\ \bibinfo {author}
  {\bibfnamefont {D.~R.}\ \bibnamefont {Wik}},\ }\href {\doibase
  10.1103/PhysRevD.107.023009} {\bibfield  {journal} {\bibinfo  {journal}
  {Phys. Rev. D}\ }\textbf {\bibinfo {volume} {107}},\ \bibinfo {pages}
  {023009} (\bibinfo {year} {2023})},\ \Eprint
  {http://arxiv.org/abs/2207.04572} {arXiv:2207.04572 [astro-ph.HE]}
  \BibitemShut {NoStop}%
\bibitem [{\citenamefont {Calore}\ \emph {et~al.}(2023)\citenamefont {Calore},
  \citenamefont {Dekker}, \citenamefont {Serpico},\ and\ \citenamefont
  {Siegert}}]{Calore:2022pks}%
  \BibitemOpen
  \bibfield  {author} {\bibinfo {author} {\bibfnamefont {F.}~\bibnamefont
  {Calore}}, \bibinfo {author} {\bibfnamefont {A.}~\bibnamefont {Dekker}},
  \bibinfo {author} {\bibfnamefont {P.~D.}\ \bibnamefont {Serpico}}, \ and\
  \bibinfo {author} {\bibfnamefont {T.}~\bibnamefont {Siegert}},\ }\href
  {\doibase 10.1093/mnras/stad457} {\bibfield  {journal} {\bibinfo  {journal}
  {Mon. Not. Roy. Astron. Soc.}\ }\textbf {\bibinfo {volume} {520}},\ \bibinfo
  {pages} {4167} (\bibinfo {year} {2023})},\ \Eprint
  {http://arxiv.org/abs/2209.06299} {arXiv:2209.06299 [hep-ph]} \BibitemShut
  {NoStop}%
\bibitem [{\citenamefont {Yin}\ \emph {et~al.}(2025{\natexlab{a}})\citenamefont
  {Yin}, \citenamefont {Fujita}, \citenamefont {Ezoe},\ and\ \citenamefont
  {Ishisaki}}]{Yin:2025xad}%
  \BibitemOpen
  \bibfield  {author} {\bibinfo {author} {\bibfnamefont {W.}~\bibnamefont
  {Yin}}, \bibinfo {author} {\bibfnamefont {Y.}~\bibnamefont {Fujita}},
  \bibinfo {author} {\bibfnamefont {Y.}~\bibnamefont {Ezoe}}, \ and\ \bibinfo
  {author} {\bibfnamefont {Y.}~\bibnamefont {Ishisaki}},\ }\href@noop {} {\
  (\bibinfo {year} {2025}{\natexlab{a}})},\ \Eprint
  {http://arxiv.org/abs/2503.04726} {arXiv:2503.04726 [hep-ph]} \BibitemShut
  {NoStop}%
\bibitem [{\citenamefont {Essig}\ \emph {et~al.}(2013)\citenamefont {Essig},
  \citenamefont {Kuflik}, \citenamefont {McDermott}, \citenamefont {Volansky},\
  and\ \citenamefont {Zurek}}]{Essig:2013goa}%
  \BibitemOpen
  \bibfield  {author} {\bibinfo {author} {\bibfnamefont {R.}~\bibnamefont
  {Essig}}, \bibinfo {author} {\bibfnamefont {E.}~\bibnamefont {Kuflik}},
  \bibinfo {author} {\bibfnamefont {S.~D.}\ \bibnamefont {McDermott}}, \bibinfo
  {author} {\bibfnamefont {T.}~\bibnamefont {Volansky}}, \ and\ \bibinfo
  {author} {\bibfnamefont {K.~M.}\ \bibnamefont {Zurek}},\ }\href {\doibase
  10.1007/JHEP11(2013)193} {\bibfield  {journal} {\bibinfo  {journal} {JHEP}\
  }\textbf {\bibinfo {volume} {11}},\ \bibinfo {pages} {193} (\bibinfo {year}
  {2013})},\ \Eprint {http://arxiv.org/abs/1309.4091} {arXiv:1309.4091
  [hep-ph]} \BibitemShut {NoStop}%
\bibitem [{\citenamefont {Porras-Bedmar}\ \emph {et~al.}(2024)\citenamefont
  {Porras-Bedmar}, \citenamefont {Meyer},\ and\ \citenamefont
  {Horns}}]{Porras-Bedmar:2024uql}%
  \BibitemOpen
  \bibfield  {author} {\bibinfo {author} {\bibfnamefont {S.}~\bibnamefont
  {Porras-Bedmar}}, \bibinfo {author} {\bibfnamefont {M.}~\bibnamefont
  {Meyer}}, \ and\ \bibinfo {author} {\bibfnamefont {D.}~\bibnamefont
  {Horns}},\ }\href {\doibase 10.1103/PhysRevD.110.103501} {\bibfield
  {journal} {\bibinfo  {journal} {Phys. Rev. D}\ }\textbf {\bibinfo {volume}
  {110}},\ \bibinfo {pages} {103501} (\bibinfo {year} {2024})},\ \Eprint
  {http://arxiv.org/abs/2407.10618} {arXiv:2407.10618 [astro-ph.CO]}
  \BibitemShut {NoStop}%
\bibitem [{\citenamefont {O'Hare}(2020)}]{AxionLimits}%
  \BibitemOpen
  \bibfield  {author} {\bibinfo {author} {\bibfnamefont {C.}~\bibnamefont
  {O'Hare}},\ }\href {\doibase 10.5281/zenodo.3932430} {\enquote {\bibinfo
  {title} {cajohare/axionlimits: Axionlimits},}\ } (\bibinfo {year}
  {2020})\BibitemShut {NoStop}%
\bibitem [{\citenamefont {Yin}\ \emph {et~al.}(2025{\natexlab{b}})\citenamefont
  {Yin}, \citenamefont {Nakagawa}, \citenamefont {Murokoshi},\ and\
  \citenamefont {Hattori}}]{Yin:2024trc}%
  \BibitemOpen
  \bibfield  {author} {\bibinfo {author} {\bibfnamefont {W.}~\bibnamefont
  {Yin}}, \bibinfo {author} {\bibfnamefont {S.}~\bibnamefont {Nakagawa}},
  \bibinfo {author} {\bibfnamefont {T.}~\bibnamefont {Murokoshi}}, \ and\
  \bibinfo {author} {\bibfnamefont {M.}~\bibnamefont {Hattori}},\ }\href
  {\doibase 10.1088/1475-7516/2025/02/063} {\bibfield  {journal} {\bibinfo
  {journal} {JCAP}\ }\textbf {\bibinfo {volume} {02}},\ \bibinfo {pages} {063}
  (\bibinfo {year} {2025}{\natexlab{b}})},\ \Eprint
  {http://arxiv.org/abs/2405.10303} {arXiv:2405.10303 [hep-ph]} \BibitemShut
  {NoStop}%
\bibitem [{\citenamefont {Cyburt}\ \emph {et~al.}(2016)\citenamefont {Cyburt},
  \citenamefont {Fields}, \citenamefont {Olive},\ and\ \citenamefont
  {Yeh}}]{Cyburt:2015mya}%
  \BibitemOpen
  \bibfield  {author} {\bibinfo {author} {\bibfnamefont {R.~H.}\ \bibnamefont
  {Cyburt}}, \bibinfo {author} {\bibfnamefont {B.~D.}\ \bibnamefont {Fields}},
  \bibinfo {author} {\bibfnamefont {K.~A.}\ \bibnamefont {Olive}}, \ and\
  \bibinfo {author} {\bibfnamefont {T.-H.}\ \bibnamefont {Yeh}},\ }\href
  {\doibase 10.1103/RevModPhys.88.015004} {\bibfield  {journal} {\bibinfo
  {journal} {Rev. Mod. Phys.}\ }\textbf {\bibinfo {volume} {88}},\ \bibinfo
  {pages} {015004} (\bibinfo {year} {2016})},\ \Eprint
  {http://arxiv.org/abs/1505.01076} {arXiv:1505.01076 [astro-ph.CO]}
  \BibitemShut {NoStop}%
\bibitem [{\citenamefont {Shi}\ and\ \citenamefont {Sigl}(1994)}]{Shi:1993ee}%
  \BibitemOpen
  \bibfield  {author} {\bibinfo {author} {\bibfnamefont {X.}~\bibnamefont
  {Shi}}\ and\ \bibinfo {author} {\bibfnamefont {G.}~\bibnamefont {Sigl}},\
  }\href {\doibase 10.1016/0370-2693(94)91232-7} {\bibfield  {journal}
  {\bibinfo  {journal} {Phys. Lett. B}\ }\textbf {\bibinfo {volume} {323}},\
  \bibinfo {pages} {360} (\bibinfo {year} {1994})},\ \bibinfo {note} {[Erratum:
  Phys.Lett.B 324, 516--516 (1994)]},\ \Eprint
  {http://arxiv.org/abs/hep-ph/9312247} {arXiv:hep-ph/9312247} \BibitemShut
  {NoStop}%
\bibitem [{\citenamefont {Suliga}\ \emph {et~al.}(2020)\citenamefont {Suliga},
  \citenamefont {Tamborra},\ and\ \citenamefont {Wu}}]{Suliga:2020vpz}%
  \BibitemOpen
  \bibfield  {author} {\bibinfo {author} {\bibfnamefont {A.~M.}\ \bibnamefont
  {Suliga}}, \bibinfo {author} {\bibfnamefont {I.}~\bibnamefont {Tamborra}}, \
  and\ \bibinfo {author} {\bibfnamefont {M.-R.}\ \bibnamefont {Wu}},\ }\href
  {\doibase 10.1088/1475-7516/2020/08/018} {\bibfield  {journal} {\bibinfo
  {journal} {JCAP}\ }\textbf {\bibinfo {volume} {08}},\ \bibinfo {pages} {018}
  (\bibinfo {year} {2020})},\ \Eprint {http://arxiv.org/abs/2004.11389}
  {arXiv:2004.11389 [astro-ph.HE]} \BibitemShut {NoStop}%
\bibitem [{\citenamefont {Azatov}\ and\ \citenamefont
  {Vanvlasselaer}(2021)}]{Azatov:2020ufh}%
  \BibitemOpen
  \bibfield  {author} {\bibinfo {author} {\bibfnamefont {A.}~\bibnamefont
  {Azatov}}\ and\ \bibinfo {author} {\bibfnamefont {M.}~\bibnamefont
  {Vanvlasselaer}},\ }\href {\doibase 10.1088/1475-7516/2021/01/058} {\bibfield
   {journal} {\bibinfo  {journal} {JCAP}\ }\textbf {\bibinfo {volume} {01}},\
  \bibinfo {pages} {058} (\bibinfo {year} {2021})},\ \Eprint
  {http://arxiv.org/abs/2010.02590} {arXiv:2010.02590 [hep-ph]} \BibitemShut
  {NoStop}%
\bibitem [{\citenamefont {Azatov}\ \emph {et~al.}(2021)\citenamefont {Azatov},
  \citenamefont {Vanvlasselaer},\ and\ \citenamefont {Yin}}]{Azatov:2021ifm}%
  \BibitemOpen
  \bibfield  {author} {\bibinfo {author} {\bibfnamefont {A.}~\bibnamefont
  {Azatov}}, \bibinfo {author} {\bibfnamefont {M.}~\bibnamefont
  {Vanvlasselaer}}, \ and\ \bibinfo {author} {\bibfnamefont {W.}~\bibnamefont
  {Yin}},\ }\href {\doibase 10.1007/JHEP03(2021)288} {\bibfield  {journal}
  {\bibinfo  {journal} {JHEP}\ }\textbf {\bibinfo {volume} {03}},\ \bibinfo
  {pages} {288} (\bibinfo {year} {2021})},\ \Eprint
  {http://arxiv.org/abs/2101.05721} {arXiv:2101.05721 [hep-ph]} \BibitemShut
  {NoStop}%
\bibitem [{\citenamefont {Baldes}\ \emph {et~al.}(2023)\citenamefont {Baldes},
  \citenamefont {Gouttenoire},\ and\ \citenamefont {Sala}}]{Baldes:2022oev}%
  \BibitemOpen
  \bibfield  {author} {\bibinfo {author} {\bibfnamefont {I.}~\bibnamefont
  {Baldes}}, \bibinfo {author} {\bibfnamefont {Y.}~\bibnamefont {Gouttenoire}},
  \ and\ \bibinfo {author} {\bibfnamefont {F.}~\bibnamefont {Sala}},\ }\href
  {\doibase 10.21468/SciPostPhys.14.3.033} {\bibfield  {journal} {\bibinfo
  {journal} {SciPost Phys.}\ }\textbf {\bibinfo {volume} {14}},\ \bibinfo
  {pages} {033} (\bibinfo {year} {2023})},\ \Eprint
  {http://arxiv.org/abs/2207.05096} {arXiv:2207.05096 [hep-ph]} \BibitemShut
  {NoStop}%
\bibitem [{\citenamefont {Azatov}\ \emph {et~al.}(2024)\citenamefont {Azatov},
  \citenamefont {Nagels}, \citenamefont {Vanvlasselaer},\ and\ \citenamefont
  {Yin}}]{Azatov:2024crd}%
  \BibitemOpen
  \bibfield  {author} {\bibinfo {author} {\bibfnamefont {A.}~\bibnamefont
  {Azatov}}, \bibinfo {author} {\bibfnamefont {X.}~\bibnamefont {Nagels}},
  \bibinfo {author} {\bibfnamefont {M.}~\bibnamefont {Vanvlasselaer}}, \ and\
  \bibinfo {author} {\bibfnamefont {W.}~\bibnamefont {Yin}},\ }\href {\doibase
  10.1007/JHEP11(2024)129} {\bibfield  {journal} {\bibinfo  {journal} {JHEP}\
  }\textbf {\bibinfo {volume} {11}},\ \bibinfo {pages} {129} (\bibinfo {year}
  {2024})},\ \Eprint {http://arxiv.org/abs/2406.12554} {arXiv:2406.12554
  [hep-ph]} \BibitemShut {NoStop}%
\bibitem [{\citenamefont {Ai}\ \emph {et~al.}(2024)\citenamefont {Ai},
  \citenamefont {Fairbairn}, \citenamefont {Mimasu},\ and\ \citenamefont
  {You}}]{Ai:2024ikj}%
  \BibitemOpen
  \bibfield  {author} {\bibinfo {author} {\bibfnamefont {W.-Y.}\ \bibnamefont
  {Ai}}, \bibinfo {author} {\bibfnamefont {M.}~\bibnamefont {Fairbairn}},
  \bibinfo {author} {\bibfnamefont {K.}~\bibnamefont {Mimasu}}, \ and\ \bibinfo
  {author} {\bibfnamefont {T.}~\bibnamefont {You}},\ }\href@noop {} {\
  (\bibinfo {year} {2024})},\ \Eprint {http://arxiv.org/abs/2406.20051}
  {arXiv:2406.20051 [hep-ph]} \BibitemShut {NoStop}%
\bibitem [{\citenamefont {Aghanim}\ \emph {et~al.}(2020)\citenamefont {Aghanim}
  \emph {et~al.}}]{Planck:2018vyg}%
  \BibitemOpen
  \bibfield  {author} {\bibinfo {author} {\bibfnamefont {N.}~\bibnamefont
  {Aghanim}} \emph {et~al.} (\bibinfo {collaboration} {Planck}),\ }\href
  {\doibase 10.1051/0004-6361/201833910} {\bibfield  {journal} {\bibinfo
  {journal} {Astron. Astrophys.}\ }\textbf {\bibinfo {volume} {641}},\ \bibinfo
  {pages} {A6} (\bibinfo {year} {2020})},\ \bibinfo {note} {[Erratum:
  Astron.Astrophys. 652, C4 (2021)]},\ \Eprint
  {http://arxiv.org/abs/1807.06209} {arXiv:1807.06209 [astro-ph.CO]}
  \BibitemShut {NoStop}%
\bibitem [{\citenamefont {Abazajian}\ \emph {et~al.}(2016)\citenamefont
  {Abazajian} \emph {et~al.}}]{CMB-S4:2016ple}%
  \BibitemOpen
  \bibfield  {author} {\bibinfo {author} {\bibfnamefont {K.~N.}\ \bibnamefont
  {Abazajian}} \emph {et~al.} (\bibinfo {collaboration} {CMB-S4}),\ }\href@noop
  {} {\  (\bibinfo {year} {2016})},\ \Eprint {http://arxiv.org/abs/1610.02743}
  {arXiv:1610.02743 [astro-ph.CO]} \BibitemShut {NoStop}%
\bibitem [{\citenamefont {Navarro}\ \emph {et~al.}(1997)\citenamefont
  {Navarro}, \citenamefont {Frenk},\ and\ \citenamefont
  {White}}]{Navarro:1996gj}%
  \BibitemOpen
  \bibfield  {author} {\bibinfo {author} {\bibfnamefont {J.~F.}\ \bibnamefont
  {Navarro}}, \bibinfo {author} {\bibfnamefont {C.~S.}\ \bibnamefont {Frenk}},
  \ and\ \bibinfo {author} {\bibfnamefont {S.~D.~M.}\ \bibnamefont {White}},\
  }\href {\doibase 10.1086/304888} {\bibfield  {journal} {\bibinfo  {journal}
  {Astrophys. J.}\ }\textbf {\bibinfo {volume} {490}},\ \bibinfo {pages} {493}
  (\bibinfo {year} {1997})},\ \Eprint {http://arxiv.org/abs/astro-ph/9611107}
  {arXiv:astro-ph/9611107} \BibitemShut {NoStop}%
\bibitem [{\citenamefont {Cautun}\ \emph {et~al.}(2020)\citenamefont {Cautun},
  \citenamefont {Benitez-Llambay}, \citenamefont {Deason}, \citenamefont
  {Frenk}, \citenamefont {Fattahi}, \citenamefont {G\'omez}, \citenamefont
  {Grand}, \citenamefont {Oman}, \citenamefont {Navarro},\ and\ \citenamefont
  {Simpson}}]{Cautun:2019eaf}%
  \BibitemOpen
  \bibfield  {author} {\bibinfo {author} {\bibfnamefont {M.}~\bibnamefont
  {Cautun}}, \bibinfo {author} {\bibfnamefont {A.}~\bibnamefont
  {Benitez-Llambay}}, \bibinfo {author} {\bibfnamefont {A.~J.}\ \bibnamefont
  {Deason}}, \bibinfo {author} {\bibfnamefont {C.~S.}\ \bibnamefont {Frenk}},
  \bibinfo {author} {\bibfnamefont {A.}~\bibnamefont {Fattahi}}, \bibinfo
  {author} {\bibfnamefont {F.~A.}\ \bibnamefont {G\'omez}}, \bibinfo {author}
  {\bibfnamefont {R.~J.~J.}\ \bibnamefont {Grand}}, \bibinfo {author}
  {\bibfnamefont {K.~A.}\ \bibnamefont {Oman}}, \bibinfo {author}
  {\bibfnamefont {J.~F.}\ \bibnamefont {Navarro}}, \ and\ \bibinfo {author}
  {\bibfnamefont {C.~M.}\ \bibnamefont {Simpson}},\ }\href {\doibase
  10.1093/mnras/staa1017} {\bibfield  {journal} {\bibinfo  {journal} {Mon. Not.
  Roy. Astron. Soc.}\ }\textbf {\bibinfo {volume} {494}},\ \bibinfo {pages}
  {4291} (\bibinfo {year} {2020})},\ \Eprint {http://arxiv.org/abs/1911.04557}
  {arXiv:1911.04557 [astro-ph.GA]} \BibitemShut {NoStop}%
\bibitem [{\citenamefont {Enqvist}\ \emph {et~al.}(2020)\citenamefont
  {Enqvist}, \citenamefont {Nadathur}, \citenamefont {Sekiguchi},\ and\
  \citenamefont {Takahashi}}]{Enqvist:2019tsa}%
  \BibitemOpen
  \bibfield  {author} {\bibinfo {author} {\bibfnamefont {K.}~\bibnamefont
  {Enqvist}}, \bibinfo {author} {\bibfnamefont {S.}~\bibnamefont {Nadathur}},
  \bibinfo {author} {\bibfnamefont {T.}~\bibnamefont {Sekiguchi}}, \ and\
  \bibinfo {author} {\bibfnamefont {T.}~\bibnamefont {Takahashi}},\ }\href
  {\doibase 10.1088/1475-7516/2020/04/015} {\bibfield  {journal} {\bibinfo
  {journal} {JCAP}\ }\textbf {\bibinfo {volume} {04}},\ \bibinfo {pages} {015}
  (\bibinfo {year} {2020})},\ \Eprint {http://arxiv.org/abs/1906.09112}
  {arXiv:1906.09112 [astro-ph.CO]} \BibitemShut {NoStop}%
\bibitem [{\citenamefont {Alvi}\ \emph {et~al.}(2022)\citenamefont {Alvi},
  \citenamefont {Brinckmann}, \citenamefont {Gerbino}, \citenamefont
  {Lattanzi},\ and\ \citenamefont {Pagano}}]{Alvi:2022aam}%
  \BibitemOpen
  \bibfield  {author} {\bibinfo {author} {\bibfnamefont {S.}~\bibnamefont
  {Alvi}}, \bibinfo {author} {\bibfnamefont {T.}~\bibnamefont {Brinckmann}},
  \bibinfo {author} {\bibfnamefont {M.}~\bibnamefont {Gerbino}}, \bibinfo
  {author} {\bibfnamefont {M.}~\bibnamefont {Lattanzi}}, \ and\ \bibinfo
  {author} {\bibfnamefont {L.}~\bibnamefont {Pagano}},\ }\href {\doibase
  10.1088/1475-7516/2022/11/015} {\bibfield  {journal} {\bibinfo  {journal}
  {JCAP}\ }\textbf {\bibinfo {volume} {11}},\ \bibinfo {pages} {015} (\bibinfo
  {year} {2022})},\ \Eprint {http://arxiv.org/abs/2205.05636} {arXiv:2205.05636
  [astro-ph.CO]} \BibitemShut {NoStop}%
\bibitem [{\citenamefont {Batell}\ and\ \citenamefont
  {Yin}(2024)}]{Batell:2024hzo}%
  \BibitemOpen
  \bibfield  {author} {\bibinfo {author} {\bibfnamefont {B.}~\bibnamefont
  {Batell}}\ and\ \bibinfo {author} {\bibfnamefont {W.}~\bibnamefont {Yin}},\
  }\href {\doibase 10.1103/PhysRevD.110.075038} {\bibfield  {journal} {\bibinfo
   {journal} {Phys. Rev. D}\ }\textbf {\bibinfo {volume} {110}},\ \bibinfo
  {pages} {075038} (\bibinfo {year} {2024})},\ \Eprint
  {http://arxiv.org/abs/2406.17028} {arXiv:2406.17028 [hep-ph]} \BibitemShut
  {NoStop}%
\bibitem [{\citenamefont {Abbasi}\ \emph {et~al.}(2022)\citenamefont {Abbasi}
  \emph {et~al.}}]{Abbasi:2021qfz}%
  \BibitemOpen
  \bibfield  {author} {\bibinfo {author} {\bibfnamefont {R.}~\bibnamefont
  {Abbasi}} \emph {et~al.},\ }\href {\doibase 10.3847/1538-4357/ac4d29}
  {\bibfield  {journal} {\bibinfo  {journal} {Astrophys. J.}\ }\textbf
  {\bibinfo {volume} {928}},\ \bibinfo {pages} {50} (\bibinfo {year} {2022})},\
  \Eprint {http://arxiv.org/abs/2111.10299} {arXiv:2111.10299 [astro-ph.HE]}
  \BibitemShut {NoStop}%
\bibitem [{\citenamefont {Aartsen}\ \emph {et~al.}(2021)\citenamefont {Aartsen}
  \emph {et~al.}}]{IceCube:2021rpz}%
  \BibitemOpen
  \bibfield  {author} {\bibinfo {author} {\bibfnamefont {M.~G.}\ \bibnamefont
  {Aartsen}} \emph {et~al.} (\bibinfo {collaboration} {IceCube}),\ }\href
  {\doibase 10.1038/s41586-021-03256-1} {\bibfield  {journal} {\bibinfo
  {journal} {Nature}\ }\textbf {\bibinfo {volume} {591}},\ \bibinfo {pages}
  {220} (\bibinfo {year} {2021})},\ \bibinfo {note} {[Erratum: Nature 592, E11
  (2021)]},\ \Eprint {http://arxiv.org/abs/2110.15051} {arXiv:2110.15051
  [hep-ex]} \BibitemShut {NoStop}%
\bibitem [{\citenamefont {Abdul~Halim}\ \emph {et~al.}(2023)\citenamefont
  {Abdul~Halim} \emph {et~al.}}]{PierreAuger:2023pjg}%
  \BibitemOpen
  \bibfield  {author} {\bibinfo {author} {\bibfnamefont {A.}~\bibnamefont
  {Abdul~Halim}} \emph {et~al.} (\bibinfo {collaboration} {Pierre Auger}),\
  }\href {\doibase 10.22323/1.444.1488} {\bibfield  {journal} {\bibinfo
  {journal} {PoS}\ }\textbf {\bibinfo {volume} {ICRC2023}},\ \bibinfo {pages}
  {1488} (\bibinfo {year} {2023})}\BibitemShut {NoStop}%
\bibitem [{\citenamefont {Albert}\ \emph {et~al.}(2024)\citenamefont {Albert}
  \emph {et~al.}}]{ANTARES:2024ihw}%
  \BibitemOpen
  \bibfield  {author} {\bibinfo {author} {\bibfnamefont {A.}~\bibnamefont
  {Albert}} \emph {et~al.} (\bibinfo {collaboration} {ANTARES}),\ }\href
  {\doibase 10.1088/1475-7516/2024/08/038} {\bibfield  {journal} {\bibinfo
  {journal} {JCAP}\ }\textbf {\bibinfo {volume} {08}},\ \bibinfo {pages} {038}
  (\bibinfo {year} {2024})},\ \Eprint {http://arxiv.org/abs/2407.00328}
  {arXiv:2407.00328 [astro-ph.HE]} \BibitemShut {NoStop}%
\bibitem [{\citenamefont {Ho}\ \emph {et~al.}(2019)\citenamefont {Ho},
  \citenamefont {Takahashi},\ and\ \citenamefont {Yin}}]{Ho:2019ayl}%
  \BibitemOpen
  \bibfield  {author} {\bibinfo {author} {\bibfnamefont {S.-Y.}\ \bibnamefont
  {Ho}}, \bibinfo {author} {\bibfnamefont {F.}~\bibnamefont {Takahashi}}, \
  and\ \bibinfo {author} {\bibfnamefont {W.}~\bibnamefont {Yin}},\ }\href
  {\doibase 10.1007/JHEP04(2019)149} {\bibfield  {journal} {\bibinfo  {journal}
  {JHEP}\ }\textbf {\bibinfo {volume} {04}},\ \bibinfo {pages} {149} (\bibinfo
  {year} {2019})},\ \Eprint {http://arxiv.org/abs/1901.01240} {arXiv:1901.01240
  [hep-ph]} \BibitemShut {NoStop}%
\bibitem [{\citenamefont {Jaeckel}\ and\ \citenamefont
  {Yin}(2021{\natexlab{b}})}]{Jaeckel:2021gah}%
  \BibitemOpen
  \bibfield  {author} {\bibinfo {author} {\bibfnamefont {J.}~\bibnamefont
  {Jaeckel}}\ and\ \bibinfo {author} {\bibfnamefont {W.}~\bibnamefont {Yin}},\
  }\href {\doibase 10.1103/PhysRevD.103.115019} {\bibfield  {journal} {\bibinfo
   {journal} {Phys. Rev. D}\ }\textbf {\bibinfo {volume} {103}},\ \bibinfo
  {pages} {115019} (\bibinfo {year} {2021}{\natexlab{b}})},\ \Eprint
  {http://arxiv.org/abs/2102.00006} {arXiv:2102.00006 [hep-ph]} \BibitemShut
  {NoStop}%
\bibitem [{\citenamefont {Yin}(2019)}]{Yin:2018yjn}%
  \BibitemOpen
  \bibfield  {author} {\bibinfo {author} {\bibfnamefont {W.}~\bibnamefont
  {Yin}},\ }\href {\doibase 10.1051/epjconf/201920804003} {\bibfield  {journal}
  {\bibinfo  {journal} {EPJ Web Conf.}\ }\textbf {\bibinfo {volume} {208}},\
  \bibinfo {pages} {04003} (\bibinfo {year} {2019})},\ \Eprint
  {http://arxiv.org/abs/1809.08610} {arXiv:1809.08610 [hep-ph]} \BibitemShut
  {NoStop}%
\bibitem [{\citenamefont {Bringmann}\ and\ \citenamefont
  {Pospelov}(2019)}]{Bringmann:2018cvk}%
  \BibitemOpen
  \bibfield  {author} {\bibinfo {author} {\bibfnamefont {T.}~\bibnamefont
  {Bringmann}}\ and\ \bibinfo {author} {\bibfnamefont {M.}~\bibnamefont
  {Pospelov}},\ }\href {\doibase 10.1103/PhysRevLett.122.171801} {\bibfield
  {journal} {\bibinfo  {journal} {Phys. Rev. Lett.}\ }\textbf {\bibinfo
  {volume} {122}},\ \bibinfo {pages} {171801} (\bibinfo {year} {2019})},\
  \Eprint {http://arxiv.org/abs/1810.10543} {arXiv:1810.10543 [hep-ph]}
  \BibitemShut {NoStop}%
\bibitem [{\citenamefont {Ema}\ \emph {et~al.}(2019)\citenamefont {Ema},
  \citenamefont {Sala},\ and\ \citenamefont {Sato}}]{Ema:2018bih}%
  \BibitemOpen
  \bibfield  {author} {\bibinfo {author} {\bibfnamefont {Y.}~\bibnamefont
  {Ema}}, \bibinfo {author} {\bibfnamefont {F.}~\bibnamefont {Sala}}, \ and\
  \bibinfo {author} {\bibfnamefont {R.}~\bibnamefont {Sato}},\ }\href {\doibase
  10.1103/PhysRevLett.122.181802} {\bibfield  {journal} {\bibinfo  {journal}
  {Phys. Rev. Lett.}\ }\textbf {\bibinfo {volume} {122}},\ \bibinfo {pages}
  {181802} (\bibinfo {year} {2019})},\ \Eprint
  {http://arxiv.org/abs/1811.00520} {arXiv:1811.00520 [hep-ph]} \BibitemShut
  {NoStop}%
\bibitem [{\citenamefont {Cappiello}\ \emph {et~al.}(2019)\citenamefont
  {Cappiello}, \citenamefont {Ng},\ and\ \citenamefont
  {Beacom}}]{Cappiello:2018hsu}%
  \BibitemOpen
  \bibfield  {author} {\bibinfo {author} {\bibfnamefont {C.~V.}\ \bibnamefont
  {Cappiello}}, \bibinfo {author} {\bibfnamefont {K.~C.~Y.}\ \bibnamefont
  {Ng}}, \ and\ \bibinfo {author} {\bibfnamefont {J.~F.}\ \bibnamefont
  {Beacom}},\ }\href {\doibase 10.1103/PhysRevD.99.063004} {\bibfield
  {journal} {\bibinfo  {journal} {Phys. Rev. D}\ }\textbf {\bibinfo {volume}
  {99}},\ \bibinfo {pages} {063004} (\bibinfo {year} {2019})},\ \Eprint
  {http://arxiv.org/abs/1810.07705} {arXiv:1810.07705 [hep-ph]} \BibitemShut
  {NoStop}%
\end{thebibliography}%

\end{document}